\documentclass[aps,onecolumn,prd,preprintnumbers,superscriptaddress,nofootinbib,bibnotes,floatfix,
longbibliography,
]{revtex4-2}

\pdfoutput=1
\usepackage{amsmath}
\usepackage{amsfonts}
\usepackage{amssymb}
\usepackage{mathrsfs}
\usepackage{graphicx}
\usepackage{color}
\usepackage{bbding}
\usepackage{tabularx}
\usepackage[usenames,dvipsnames,table]{xcolor}
\usepackage[normalem]{ulem}
\usepackage{makecell}
\usepackage{siunitx}
\usepackage{enumerate}
\usepackage{multirow}
\usepackage{makecell}
\usepackage{upgreek}
\usepackage{comment}
\usepackage{framed}
\usepackage{subfigure}
\usepackage{booktabs}

\usepackage{marvosym}
\definecolor{darkred}{rgb}{0.5,0,0}
\definecolor{darkblue}{rgb}{0,0,0.5}
\definecolor{firebrick}{rgb}{0.75,0.125,0.125}
\definecolor{darkgreen}{rgb}{0,0.5,0}
\usepackage[colorlinks=true,linkcolor=firebrick,citecolor=darkgreen,urlcolor=darkblue]{hyperref}

\newcommand{\beq}{\begin{equation}}
\newcommand{\eeq}{\end{equation}}

\newcommand{\orcid}[1]{\href{https://orcid.org/#1}{\includegraphics[width=10pt]{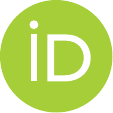}}}

\begin{document}

\title{Hubble Tension as an Effect of Horizon Entanglement Nonequilibrium}

\author{Alexander S.~Sakharov \orcid{0000-0001-6622-2923}}
\email{alexandre.sakharov@cern.ch}
\affiliation{Department of Mathematics and Physics, Manhattan University,\\
4513 Manhattan College Parkway, Riverdale, NY 10471, United States of America}
\affiliation{Experimental Physics Department, CERN, CH-1211 Gen\`eve 23, Switzerland}

\author{Rostislav Konoplich \orcid{0000-0002-6223-7017}}
\email{rostislav.konoplich@manhattan.edu}
\affiliation{Department of Mathematics and Physics, Manhattan University,\\
4513 Manhattan College Parkway, Riverdale, NY 10471, United States of America}
\affiliation{Department of Physics, New York University,\\
726 Broadway, New York, NY 10003, United States of America}

\author{Merab Gogberashvili \orcid{0000-0002-7399-0813}}
\email{gogber@gmail.com}
\affiliation{Faculty of Exact and Natural Sciences, Javakhishvili Tbilisi State University,
Tbilisi 0179, Georgia}
\affiliation{Department of Elementary Particles, Andronikashvili Institute of Physics,
Tbilisi 0177, Georgia}

\author{Jack Simoni 
}
\email{jsimoni02@manhattan.edu}
\affiliation{Department of Mathematics and Physics, Manhattan University,\\
4513 Manhattan College Parkway, Riverdale, NY 10471, United States of America}

\date{\today}

\begin{abstract}
We propose an infrared mechanism for alleviating the Hubble constant tension, based on a small departure from entanglement equilibrium at the cosmological apparent horizon.
If the horizon entanglement entropy falls slightly below the
Bekenstein--Hawking value, we parametrize the shortfall by a fractional deficit
$\delta(a)$ evolving with the FLRW scale factor $a$. The associated
equipartition deficit at the Gibbons--Hawking temperature then sources a smooth,
homogeneous component whose density scales as $H^{2}/G$, with a dimensionless
coefficient $c_{e}^{2}(a)$ of order unity times $\delta(a)$.
Because this component tracks $H^{2}$, it is negligible at early times but can activate at redshifts $z\lesssim 1$, raising the late time expansion rate by a few percent without affecting recombination or the sound horizon. We present a minimal three parameter activation model for $c_{e}^{2}(a)$ and derive its impact on the background expansion, effective equation of state, and linear growth for a smooth entanglement sector. The framework
predicts a small boost in $H(z)$, a mild suppression of $f\sigma_{8}(z)$, and a
corresponding modification of the low--$z$ distance--redshift relation. We test
these predictions against current low--redshift data sets, including SN~Ia
distance moduli, baryon acoustic oscillation distance measurements,
cosmic chronometer $H(z)$ data, and redshift space distortion
constraints, and discuss whether the $H_0$ tension can be consistently
interpreted as a late--time, horizon--scale information deficit rather than an
early universe modification.
\end{abstract}

\maketitle


\section{Introduction}
\label{sec:intro}

A persistent discrepancy has emerged between the Hubble constant inferred from
early universe probes and the value obtained from late-time distance ladders,
a percent level tension that, if not due to unrecognized systematics, hints at
new physics in the cosmic expansion history. Concretely, the CMB anisotropy
spectrum tightly constrains the angular acoustic scale and, within a specified
background model, translates this into an inferred $H_0$ through the sound
horizon and the late--time distance to last scattering. For example,
\emph{Planck} 2018 (under $\Lambda$CDM) finds
$H_0 = 67.4\pm 0.5\ \mathrm{km\,s^{-1}\,Mpc^{-1}}$~\cite{Planck2018}, whereas
local distance ladder calibrations anchored by geometric and stellar
standardization methods yield higher values; in particular, SH0ES reports
$H_0 = 73.30\pm 1.04\ \mathrm{km\,s^{-1}\,Mpc^{-1}}$~\cite{Riess2022}.
At intermediate redshifts, baryon acoustic oscillations (BAO) provide a robust
standard ruler that strongly constrains the late-time distance-redshift
relation; recent DESI BAO constraints favor values closer to the
\emph{Planck} inferred scale~\cite{DESI2024VI_BAO,DESI2025DR2_BAO}, sharpening
the question of whether a consistent cosmological history can reconcile early-
and late-time determinations.
Explanations proposed so far broadly fall into two classes. Early time
modifications (e.g., early dark energy, extra radiation, or altered
recombination) reduce the sound horizon so the CMB can accommodate a larger
$H_0$, while late-time ideas (e.g., evolving dark energy or modified gravity)
adjust the low-$z$ Hubble rate with minimal impact on recombination; see
Refs.~\cite{Vagnozzi:2019Iconsist,DiValentino2021Review,Abdalla2022Intertwined,Schoneberg2021qvd,Vagnozzi:7_hints,CosmoVerseNetwork:2025alb,Vagnozzi:2025BAO}
for reviews. Both directions face tight joint constraints from CMB anisotropies, BAO,
Type-Ia supernova distances, and structure growth observables: DESI BAO remain
consistent with flat $\Lambda$CDM and bound $w$ near $-1$, Pantheon+ tightens the
low-$z$ distance--redshift relation, and redshift distortion (RSD) growth measurements strongly
restrict early time solutions such as EDE when combined with CMB data~\cite{DESI2024VI_BAO,Scolnic2022PantheonPlusData,1607.03155,Alam2021eBOSS,DESI2025DR2_BAO,Ivanov2020EDE,Hill2022ACTEDE,Vagnozzi:7_hints}.

In parallel, advances in quantum information and gravity have reshaped how we think about spacetime: horizons carry entropy and temperature; entanglement can encode geometry; and thermodynamic reasoning can reproduce gravitational dynamics.
However, naive free-field entanglement in curved spacetime contributes only $\mathcal{O}(H^4)$ to the vacuum energy far too small to matter cosmologically and the ultra violet (UV) area law renormalizes $1/G$ rather than producing a new fluid.
Any entanglement driven mechanism relevant to the Hubble tension must therefore be infrared (IR) and horizon tied, not a re-counting of UV modes.

We develop and assess such an IR mechanism: the \emph{horizon entanglement equipartition deficit} (HEED).
The core hypothesis is that the quantum state of our late time universe is slightly out of entanglement equilibrium at the apparent (Hubble) horizon.
If the actual horizon entanglement entropy $S_{\rm ent}$ falls short of the Bekenstein--Hawking value $S_{\rm BH}=A/(4G)$ by a fractional deficit $\delta(a)\in[0,1]$, then treating the horizon as an entanglement screen at the Gibbons--Hawking temperature $T_{\rm dS}=H/(2\pi)$ the resulting equipartition deficit translates into a homogeneous bulk component of density
\begin{equation}
\rho_{\rm HEED}(a) = \frac{3}{8\pi G}\,c_e^2(a)\,H^2(a),\qquad
c_e^2(a) \sim \mathcal{O}(1)\times \delta(a),
\label{eq:intro_rho_HEED}
\end{equation}
where \(H(a)\) stand for the Hubble rate evolving with the FLRW scale factor  \(a\).
This is the familiar holographic $H^2/G$~\cite{Cohen1999,Li2004,Hsu2004,WangWangLi2017}, but with a time-dependent coefficient tied to horizon entanglement rather than an ad hoc constant.

Two features make HEED cosmologically attractive.
First, because $\rho_{\rm HEED}\propto H^2$, it is automatically negligible at
high redshift relative to matter and radiation, thereby preserving recombination
physics and the sound horizon.
Second, we model a \emph{late--time activation} of the HEED amplitude by letting
\begin{equation}
c_e^2(a)=c_{e0}^2\,g(a;a_t,k),
\label{eq:intro_c_activation}
\end{equation}
where the dimensionless switch function $g(a;a_t,k)$ interpolates smoothly from
$g\!\ll\!1$ at early times ($a\!\ll\!a_t$) to $g\!\to\!1$ at late times
($a\!\gtrsim\!a_t$). The parameter $a_t$ sets the characteristic activation
epoch (equivalently $z_t\simeq a_t^{-1}-1$), while $k$ controls the sharpness of
the transition (larger $k$ corresponds to a steeper switch). This late activation
can yield a few percent increase in the low--$z$ Hubble rate, sufficient to ease
the tension while remaining consistent with SN/BAO distances. Interpreted as a
smooth, non--clustering sector (rest--frame sound speed $\simeq 1$), HEED modestly
suppresses structure growth through $f\sigma_8$,
providing an additional levers beyond purely geometric constraints.

Conceptually, HEED can be read in two equivalent ways.
In a \emph{fluid} picture, $\rho_{\rm HEED}$ adds to the stress--energy budget
and the Friedmann equation becomes
\begin{equation}
H^2(a)=\frac{8\pi G}{3}\,
\frac{\rho_m(a)+\rho_\Lambda}{1-c_e^2(a)}\,,
\label{eq:H_with_HEED}
\end{equation}
where $\rho_m(a)=\rho_{m0}\,a^{-3}$ is the (pressureless) matter density and
$\rho_\Lambda\equiv \Lambda/(8\pi G)$ is the vacuum energy density
associated with the cosmological constant. The entanglement deficit therefore
appears as a multiplicative correction to the expansion rate.
In a \emph{modified gravity} reading, the same factor reshapes the effective
Planck mass,
$M_*^2(a)=M_{\rm Pl}^2\,[1-c_e^2(a)]$, while matter remains covariantly
conserved. Either view preserves the Bianchi identity and avoids
ghost/gradient instabilities so long as the entanglement sector is treated as
smooth and non--propagating (an effective IR dressing of the background rather
than a new field).

This paper makes following contributions.
We formalize HEED from horizon thermodynamics and entanglement equilibrium, clarifying the UV/IR split: UV area law pieces renormalize $1/G$, whereas the IR deficit $\delta(a)$ sources $\rho \propto H^2/G$.
We derive background relations closed form, an instantaneous effective $w_{\rm HEED}(a)$, and the logarithmic derivative $d\ln H/d\ln a$ needed for growth under a minimal three parameter activation $\{c_{e0}^2,a_t,k\}$.
We analyze linear perturbations with a smooth entanglement sector, predicting a characteristic pattern: small late time boosts to $H(z)$ and mild suppression of $f\sigma_8$. We confront the model with low--redshift data by fitting SN~Ia, BAO, and
cosmic chronometer (CC) constraints together with a CMB distance prior that
keeps the sound horizon fixed, and by testing the resulting growth history
against RSD measurements, while imposing the early--time safety condition
$c_e^2(a_\star)\ll 1$ at recombination.

HEED sits at the intersection of holographic dark energy (HDE) type scalings and entanglement equilibrium thinking.
Unlike standard HDE with a constant $c^2$, HEED explains the coefficient as a dynamical horizon information
shortfall that can switch on at $z\lesssim 1$.
Unlike free field ``entanglement dark energy,'' it avoids the $\mathcal{O}(H^4)$ suppression by tying directly to the horizon's IR thermodynamics.

It is important to note that the present work should be viewed as an effective IR cosmological
parametrization motivated by horizon thermodynamics and entanglement equilibrium,
rather than as a complete microscopic holographic dual or a UV-complete local
field theory. Its purpose is to isolate the phenomenological consequences of a
small late-time horizon entanglement deficit, clarify the associated UV/IR split,
and test whether such a component can remain consistent with current low-redshift
distance and growth data. This effective IR status is not unusual in cosmology: many dark-energy and modified-gravity models 
were analyzed extensively at the phenomenological level before any full UV completion or microscopic derivation was available, including early dark energy, interacting dark energy, phenomenological growth frameworks, $f(R)$ gravity, emergent dark energy, modified-gravity interpretations of late-time data, effective running Hubble constant and slow-rolling scalar dynamics~\cite{Doran2007EDE,CalderaCabral2009IDE,Linder2009GrowthFramework,HuSawicki2007fR,Montani:2025nmz,SongHuSawicki2007LSSfR,Cruz2023Profiling,Herold2023Resolving,Yao2024Observational,NazariPooya2024Growth,Chudaykin2024ModifiedGravity,Escamilla:2024xmz,Luongo2024Horndeski,Adil2024Omnipotent,Fazzari:2025mww,Montani:2024ntj,Montani:2025rcy,Montani:2023ywn}.

Recent alternatives to standard EDE include positive pressure
barotropic components \cite{CarloniLuongo2026Barotropic},
transition based dynamical dark energy with AdS-to-dS or dS-to-dS evolution
\cite{Akarsu2025AdSdS}, and HDE models with different
infrared cutoffs \cite{Carloni:2025jlk}. HEED is also infrared-motivated, but it
is tied to the apparent horizon and interprets the late time correction as a
finite entanglement equipartition deficit.

The paper is organized as follows.
Section~\ref{sec:thermo-derivation} derives HEED from horizon thermodynamics and
entanglement considerations, establishes the characteristic $H^{2}/G$ scaling,
and clarifies the required UV/IR separation; it also presents the background
dynamics, the instantaneous effective equation of state, and the minimal
activation parameterization.
Section~\ref{sec:observations} confronts the HEED activation parameters with a
combination of four low--$z$ probes. The Bayesian inference framework is
described in Sec.~\ref{sec:bayes}, where the posterior is sampled with the
affine--invariant ensemble sampler \texttt{emcee}. Marginalized constraints and
triangle plots are obtained with \texttt{GetDist} and interpreted in
Sec.~\ref{sec:triangle}, which discusses posterior constraints and the internal
consistency of the HEED picture. Section~\ref{sec:conclusions} summarizes the
main results and outlines directions for future work.

\section{Horizon Entanglement Equipartition Deficit}
\label{sec:thermo-derivation}

The HEED framework starts from the observation that, in a spatially flat FLRW universe, the apparent (Hubble) horizon behaves thermodynamically: it has a well-defined area and an associated temperature and entropy, in direct analogy with de Sitter and black hole horizons~\cite{Gibbons1977,Bekenstein1973,Hawking1975,Wald1993}. We posit that the \emph{renormalized} bulk entanglement entropy of quantum fields across this horizon is slightly below the Bekenstein–Hawking value~\cite{Gibbons1977,Bekenstein1973,Hawking1975,Wald1993}. This small shortfall, which we call “entanglement deficit”, encodes a departure from entanglement equilibrium and, when translated through horizon
equipartition~\cite{Padmanabhan2010RPP,Padmanabhan2010MPLA,Verlinde2011,Padmanabhan2012}, manifests as an infrared, horizon tied contribution to the cosmic energy budget whose magnitude tracks the square of the expansion rate as $H^{2}/G$. Because this contribution is naturally tiny at early times and only becomes relevant as the universe accelerates, it can modify the late time expansion without upsetting recombination physics, unlike free field vacuum effects whose influence scales much more weakly with the Hubble rate (see Appendix~\ref{sec:H4-scaling} for details).

Let $\Sigma_H$ denote the apparent (Hubble) horizon on a constant time FLRW slice, and decompose the Cauchy slice into an inside region $\mathcal{R}_{\rm in}$ bounded by $\Sigma_H$ and its complement $\mathcal{R}_{\rm out}$. In a QFT without gauge constraints one may (formally) factorize the Hilbert space as $\mathcal{H}\simeq\mathcal{H}_{\rm in}\otimes\mathcal{H}_{\rm out}$; with gauge or gravitational constraints it is cleaner to work with local operator algebras (possibly with a shared center or edge modes on $\Sigma_H$) that reproduce the usual reduced--state construction~\cite{CasiniHuerta2009,DonnellyFreidel2016,Harlow2017}. The global (total) state $\rho_{\rm tot}$ is a positive, unit--trace operator on the full Hilbert space, $\mathrm{Tr}\,\rho_{\rm tot}=1$. Tracing out exterior degrees of freedom defines the reduced inside state
\begin{equation}
\rho_{\rm in}(a)\;\equiv\;\mathrm{Tr}_{\rm out}\,\rho_{\rm tot}(a),
\label{eq:rhoin-def}
\end{equation}
and the entanglement entropy across $\Sigma_H$ is
\begin{equation}
S_{\rm ent}(a)\;=\;-\mathrm{Tr}_{\rm in}\,\rho_{\rm in}(a)\,\ln\rho_{\rm in}(a).
\label{eq:Sdef}
\end{equation}
As in flat space, vacuum entanglement entropy across a smooth surface exhibits an area--law divergence~\cite{Bombelli1986,Srednicki1993,CasiniHuerta2009},
\begin{equation}
S_{\rm ent}^{\rm bare}\;=\;\eta\,\frac{A(\Sigma_H)}{\epsilon^{2}}+\cdots,
\end{equation}
with nonuniversal coefficient $\eta$ and cutoff $\epsilon$. In semiclassical gravity, the same UV modes that produce the area term renormalize the Einstein--Hilbert coupling, so the generalized entropy relevant for horizon thermodynamics reads
\begin{equation}
S_{\rm gen}[\Sigma_H]\;=\;\frac{A(\Sigma_H)}{4\,G_{\rm ren}}+S_{\rm bulk}^{\rm ren}[\Sigma_H]+\cdots,
\label{eq:Sgen}
\end{equation}
where the divergence has been absorbed into $G_{\rm ren}$ and $S_{\rm bulk}^{\rm ren}$ is the finite, IR renormalized bulk entropy term
associated with quantum fields
across $\Sigma_H$~\cite{Susskind1994,Solodukhin2011,CoopermanLuty2014,EngelhardtWall2015}.
In what follows, $S_{\rm ent}(a)$ denotes this renormalized IR contribution,
\begin{equation}
S_{\rm ent}(a)\;\equiv\;S_{\rm bulk}^{\rm ren}[\Sigma_H(a)],
\end{equation}
which we compare to the Bekenstein--Hawking value
$S_{\rm BH}=A(\Sigma_H)/(4\,G_{\rm ren})$.
Departures from entanglement equilibrium are then captured by the fractional deficit
\begin{equation}
\delta(a)\;\equiv\;1-\frac{S_{\rm ent}(a)}{S_{\rm BH}(a)}\in[0,1],
\label{eq:delta_def}
\end{equation}
with $\delta=0$ expected in exact de Sitter and $\delta>0$ natural out of equilibrium at late times~\cite{Jacobson2016,Lashkari2014}.  In this sense, the entanglement deficit is not introduced as an additional UV area-law contribution, which is already absorbed into $G_{\rm ren}$, but as a
finite infrared departure from horizon entanglement saturation. The proposal is
therefore intrinsically an effective late-time parametrization of nonequilibrium
horizon thermodynamics.
Using the apparent horizon temperature $T_{\rm dS}=H/(2\pi)$~\cite{Gibbons1977}, we will map this deficit to a smooth IR component scaling as $H^{2}/G$.

To construct the HEED term, we treat the
apparent horizon as an entanglement screen
with area and temperature
\begin{equation}
A=\frac{4\pi}{H^{2}},
\qquad
T_{\rm dS}=\frac{H}{2\pi},
\label{eq:area-temp}
\end{equation}
and Bekenstein--Hawking entropy $S_{\rm BH}=A/(4G)$~\cite{Gibbons1977,Bekenstein1973,Hawking1975,Wald1993}.
Calling it an \emph{entanglement screen} means we regard the horizon as the codimension one surface across which we compute the (renormalized) bulk entanglement $S_{\rm ent}$ of quantum fields. The screen then supplies a ``bookkeeping device'' for associating a surface energy via equipartition
to the bulk IR component.
This language is in line with horizon thermodynamics, holographic ideas, and (membrane paradigm)
coarse graining of horizon
dof~\cite{Padmanabhan2010RPP,Padmanabhan2010MPLA,Verlinde2011,Padmanabhan2012,Bousso2002,Membrane1986,Lee:2007zq,Lee:2010bg}.

Following the emergent–gravity/equipartition literature, we count the effective surface degrees of freedom (DoF) as\footnote{The \emph{Planck length} is $L_p\!\equiv\!\sqrt{\hbar G/c^3}$; in the natural units used here
($\hbar=c=\kappa =1$) one has $L_p^2=G$. It sets the area quantum that appears in black hole horizon thermodynamics~\cite{Bekenstein1973,Wald1993}.}
\begin{equation}
N_{\rm surf}\;=\;\frac{A}{L_{p}^{2}}
\;=\;\frac{A}{G}
\;=\;\frac{4\pi}{G\,H^{2}}
\;=\;4\,S_{\rm BH},
\label{eq:Nsurf}
\end{equation}
which follows directly from the area law for $S_{\rm BH}$~\cite{Bekenstein1973,Wald1993}. This is a coarse grained count that does not assume a specific microscopic model and is routinely employed in horizon equipartition derivations of gravitational dynamics~\cite{Padmanabhan2010MPLA,Padmanabhan2010RPP,Padmanabhan2012,Verlinde2011}. We then assume equipartition on the screen,
\begin{equation}
E_{\rm surf}\;=\;\tfrac{1}{2}\,N_{\rm surf}\,T_{\rm dS},
\end{equation}
with de Sitter temperature $T_{\rm dS}=H/(2\pi)$~\cite{Gibbons1977}. If only a fraction $1-\delta(a)$ of these DoF are entanglement active, the \emph{missing} (deficit) energy on the screen is
\begin{equation}
\Delta E_{\rm surf}
=\tfrac{1}{2}\,\delta(a)\,N_{\rm surf}\,T_{\rm dS}
=\delta(a)\,\frac{1}{G\,H},
\label{eq:Esurf_deficit}
\end{equation}
where the last equality follows from inserting $N_{\rm surf}$ in Eq.~\eqref{eq:Nsurf} and $T_{\rm dS}=H/(2\pi)$.
Associating this with a homogeneous bulk component within the Hubble volume $V_H=\tfrac{4\pi}{3}H^{-3}$ gives
\begin{equation}
\rho_{\rm HEED}(a)=\frac{\Delta E_{\rm surf}}{V_H}
=\frac{3}{4\pi}\,\delta(a)\,\frac{H^{2}(a)}{G}
=\frac{3}{8\pi G}\,c_e^{2}(a)\,H^{2}(a),
\qquad
c_e^{2}(a)\equiv 2\,\delta(a),
\label{eq:rho_heed}
\end{equation}
up to an overall numerical convention of order unity. This realizes the familiar holographic $H^{2}/G$ scaling while anchoring it to IR horizon entanglement rather than to the UV area law, whose divergence renormalizes $1/G$ and therefore does not represent a dynamical fluid~\cite{Susskind1994,Solodukhin2011,CoopermanLuty2014}.
Our statement that HEED “anchors” this scaling to IR entanglement means: instead of postulating $\rho\propto H^{2}$, we derive it from the horizon’s thermodynamics and a finite, IR entanglement deficit $S_{\rm ent}<S_{\rm BH}$, via the equipartition argument in the text. By contrast, the UV area law term in $S_{\rm ent}$ is regulator dependent and simply renormalizes $1/G$; it is not a dynamical fluid~\cite{Susskind1994,Solodukhin2011,CoopermanLuty2014}. Thus HEED ties the coefficient to a physical deficit $\delta(a)$, rather than treating $c_e^2$ as an external constant.

There are two equivalent but conceptually distinct embeddings of Eq.~\eqref{eq:rho_heed} in FLRW. In a fluid (additive) view one writes
\begin{equation}
\label{eq:Ha0}
H^{2}(a)=\frac{8\pi G}{3}\big[\rho_{m}(a)+\rho_{\Lambda}+\rho_{\rm HEED}(a)\big],
\qquad
\rho_{\rm HEED}(a)=\frac{3}{8\pi G}\,c_e^{2}(a)H^{2}(a),
\end{equation}
which leads to the relation
\begin{equation}
H^{2}(a)=\frac{8\pi G}{3}\,
\frac{\rho_{m}(a)+\rho_{\Lambda}}{1-c_e^{2}(a)}.
\label{eq:H2_algebraic}
\end{equation}
Alternatively, in a modified--gravity (multiplicative) view, one moves the factor $1-c_e^{2}(a)$ to the geometric side and interprets it as a slowly varying Planck mass~\cite{BelliniSawicki2014,KaseTsujikawa2019},
$M_{*}^{2}(a)=M_{\rm Pl}^{2}[\,1-c_e^{2}(a)\,]$.
In either interpretation the Bianchi identity is respected and matter remains covariantly conserved, $\dot\rho_{m}+3H\rho_{m}=0$~\footnote{The twice contracted Bianchi identity $\nabla_\mu G^{\mu\nu}=0$ is a geometric identity~\cite{WaldGR,CarrollBook}. With Einstein’s equation $G^{\mu\nu}=8\pi G\,T^{\mu\nu}$ this implies covariant conservation $\nabla_\mu T^{\mu\nu}=0$. For a homogeneous fluid component this yields $\dot\rho+3H(\rho+p)=0$, and for pressureless matter $\dot\rho_m+3H\rho_m=0$, as used in the text.}.
Early time safety is guaranteed by requiring $c_e^{2}(a_{\star})\ll 1$ at recombination so the sound horizon and CMB acoustic physics remain unchanged.

For phenomenology we model a smooth late--time activation as follows,
\begin{equation}
c_e^{2}(a)=c_{e0}^{2}\,g(a),
\qquad
g(a)=\frac{1+(a_t)^k}{1+(a_{t}/a)^{k}},
\qquad
g(1)=1,
\label{eq:activation}
\end{equation}
with $a_{t}\in[0.4,0.8]$ and $k\gtrsim 3$. The normalization $g(1)=1$ makes $c_{e0}^2\!=\!c_e^2(a{=}1)$ the present day HEED fraction. Equation~\eqref{eq:H2_algebraic}  rescales the Friedman equation by factor
\begin{equation}
Q^{2}(a)\equiv\frac{H^{2}(a)}{H_{0}^{2}}
=\frac{\Omega_{m}a^{-3}+\Omega_{\Lambda}}{1-c_{e0}^{2}\,g(a)}\,,\qquad
\Omega_\Lambda=1-\Omega_m\ .
\label{eq:E2}
\end{equation}
An instantaneous effective equation of state for the HEED sector follows from $\rho_{\rm HEED}\propto H^{2}c_e^{2}$,
\begin{equation}
w_{\rm HEED}(a)=-1-\frac{1}{3}\frac{d}{d\ln a}\ln\!\big[Q^{2}(a)c_e^{2}(a)\big]
=-1-\frac{1}{3}\left(\frac{d\ln Q^{2}}{d\ln a}+\frac{d\ln c_e^{2}}{d\ln a}\right),
\label{eq:wheed}
\end{equation}
with
\begin{equation}
\frac{d\ln Q^{2}}{d\ln a}
=\frac{-3\,\Omega_{m}a^{-3}}{\Omega_{m}a^{-3}+\Omega_{\Lambda}}
+\frac{1}{1-c_e^{2}(a)}\,\frac{d c_e^{2}(a)}{d\ln a}.
\label{eq:dlnE2}
\end{equation}
Once $c_e^{2}$ has saturated ($d\ln c_e^{2}/d\ln a\simeq 0$), one finds $w_{\rm HEED}\approx -1-\tfrac{1}{3}\,d\ln Q^{2}/d\ln a\approx -1+\Omega_{m}(a)$, i.e., quintessence like behavior today for $\Omega_{m0}\simeq 0.3$.
It is important to stress that this equation of state is not introduced as an
independent microphysical assumption. Once \(\rho_{\rm HEED}\) is fixed by the
horizon-equipartition ansatz, the effective pressure is determined by the
continuity equation. The HEED sector is therefore covariantly conserved at the
homogeneous level by construction. A derivation of \(c_e^2(a)\) from a
microscopic covariant action or entropy functional remains an open theoretical
problem.

During the switch on ($d\ln c_e^{2}/d\ln a>0$), a transient $w_{\rm HEED}< -1$ can occur. This phantom behavior pertains only to the background effective equation of state and, because HEED introduces no propagating field, it does not signal a ghost. A ghost is a propagating mode with a negative kinetic term, which renders the Hamiltonian unbounded. In a canonical single field, achieving $w<-1$ indeed requires the wrong sign kinetic term and is pathological~\cite{Caldwell2002}, and $k$-essence generically cannot stably cross $w=-1$ without ghost/gradient problems~\cite{Vikman2005}. By contrast, within the EFT/Horndeski framework one can realize an effective $w_{\rm eff}< -1$ while remaining ghost free provided the kinetic matrices are positive, the scalar sound speed satisfies $c_s^2>0$, and the effective Planck mass remains positive ($M_*^2>0$), e.g., in the $\alpha$-parameterization~\cite{Gubitosi2013,Bloomfield2013,BelliniSawicki2014,KaseTsujikawa2019,Kobayashi2011,Deffayet2010}. HEED belongs to this safe class: it is an effective, nonpropagating IR sector equivalent to a mild time variation of $M_*^2(a)=M_{\rm Pl}^2[1-c_e^2(a)]$ with $0\le c_e^2(a)<1$, so any transient $w_{\rm HEED}< -1$ during the switch-on of $c_e^2(a)$ is a kinematic background effect rather than a ghost instability. Taking $\alpha_T=0$ and $\alpha_B\simeq\alpha_K\simeq 0$~\footnote{${\alpha_T}$ (tensor-speed excess) controls the propagation speed of gravitational waves, $c_T^2=1+\alpha_T$. Low-redshift bounds
from GW170817/GRB170817A imply $|\alpha_T|\!\approx\!0$ today~\cite{KaseTsujikawa2019}. ${\alpha_B}$ (braiding)
quantifies kinetic mixing between the scalar sector and the metric (“kinetic gravity braiding”), modifying the Poisson equation and the metric slip and thus growth/ISW signals~\cite{Deffayet2010,BelliniSawicki2014}. ${\alpha_K}$ (kineticity) measures the normalization of the scalar perturbation’s kinetic term in unitary gauge; it affects the scalar sound speed and no-ghost/no-gradient stability conditions but does not directly modify the tensor sector~\cite{BelliniSawicki2014,Kobayashi2011}.} keeps the extra scalar nondynamical while satisfying standard no-ghost/no-gradient conditions~\cite{BelliniSawicki2014,KaseTsujikawa2019}.

\begin{figure}[t]
\centering
\begin{minipage}{0.48\linewidth}
    \includegraphics[width=\linewidth]{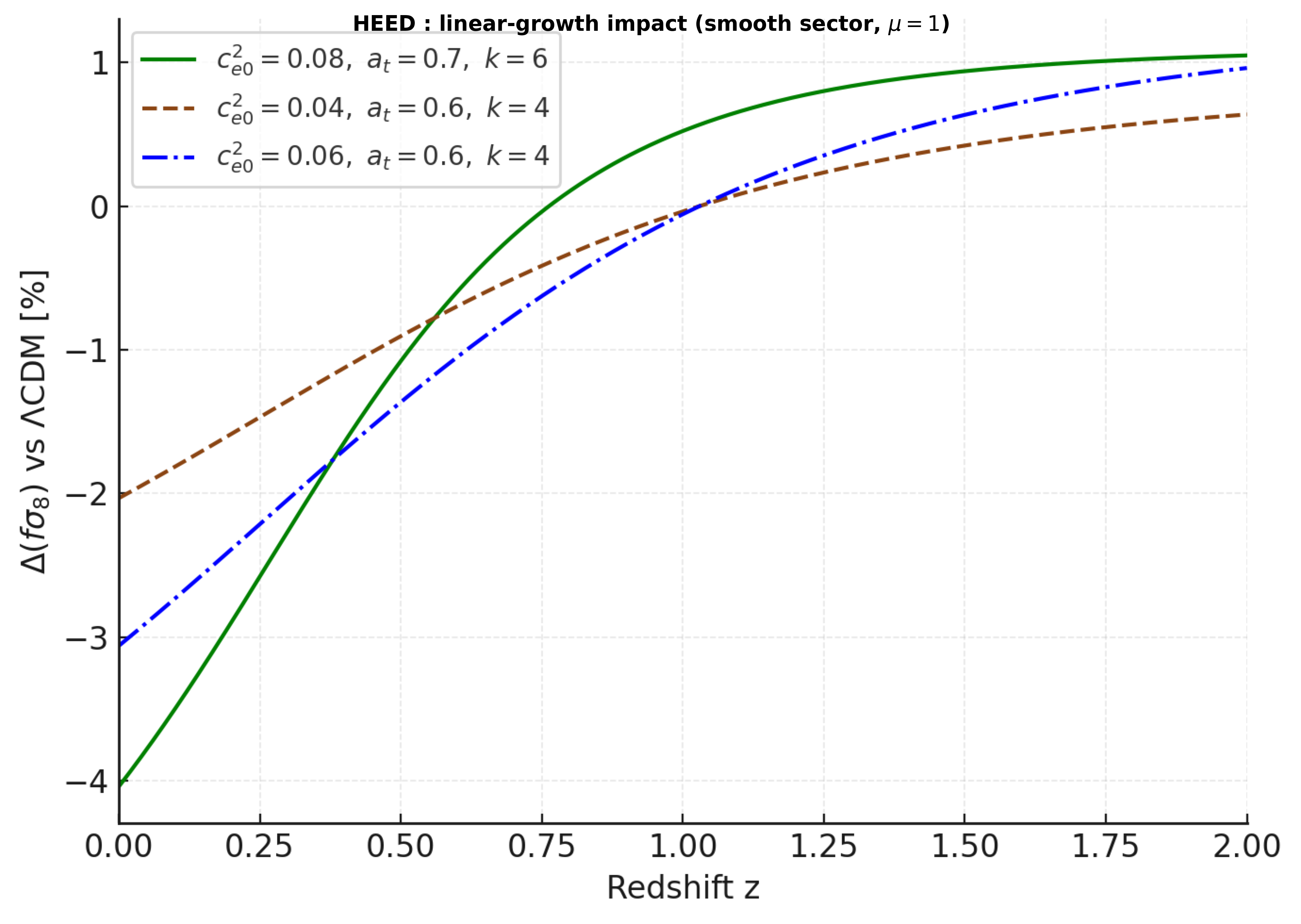}
    \caption{HEED's linear growth impact illustrated for three sets of the activation parameters.}
    \label{fig:fs8}
\end{minipage}
\hfill
\begin{minipage}{0.48\linewidth}
    \includegraphics[width=\linewidth]{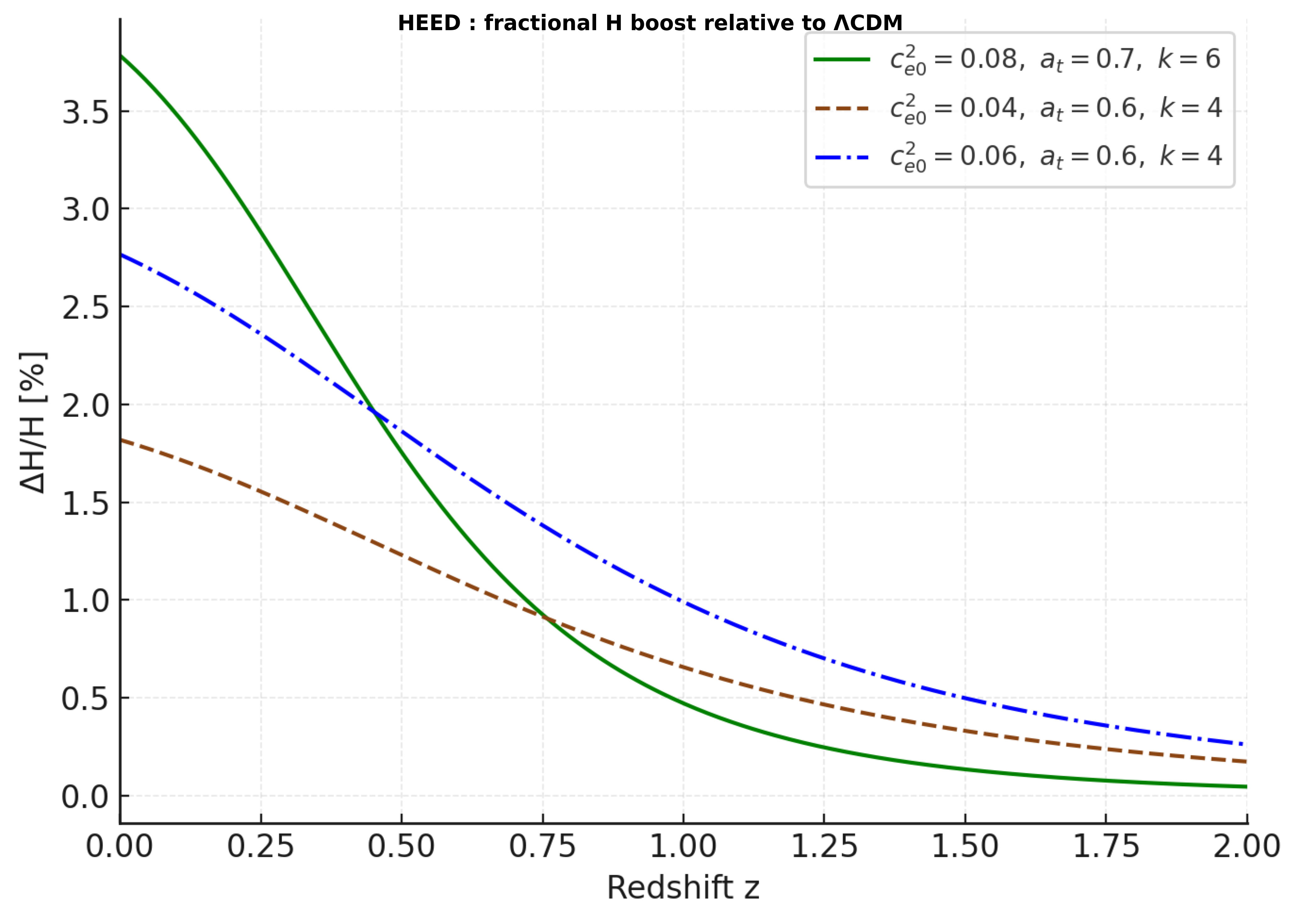}
    \caption{HEED corrected low redshift Hubble constant related via its residuals relative to the
    \(\Lambda\)CDM calculations.}
    \label{fig:Hz}
\end{minipage}
\end{figure}

In the HEED framework we treat the entanglement sector as smooth (non-clustering) and work in GR for linear perturbations. The matter growth factor $D(a)$ (normalized so that $D(1)=1$) obeys the standard ODE
\begin{equation}
D''+\Big[2+\frac{d\ln H}{d\ln a}\Big]D'-\frac{3}{2}\,\Omega_{m}(a)\,D=0,
\qquad
\Omega_{m}(a)=\frac{\Omega_{m}a^{-3}}{Q^{2}(a)},
\label{eq:heed-growth-ode}
\end{equation}
with primes denoting $d/d\ln a$ and initial conditions set deep in matter domination~\footnote{See, e.g., \cite{Peebles1980,Dodelson2003,Linder2005} for derivations and discussions of linear growth in GR.}. Defining the dimensionless logarithmic growth rate
\begin{equation}
f(a)\;\equiv\;\frac{d\ln D}{d\ln a},
\label{eq:def-f}
\end{equation}
and the fluctuation amplitude within $8\,h^{-1}{\rm Mpc}$ spheres (with $h \equiv H_0/100\,{\rm km\,s^{-1}\,Mpc^{-1}}$ the reduced Hubble constant),
\begin{equation}
\sigma_8(a)\;\equiv\;\sigma_{8,0}\,D(a),
\label{eq:def-s8}
\end{equation}
the RSD observable is the product $f\sigma_8(z)$ which is robust to linear galaxy bias and widely
used to test late time dynamics \cite{Kaiser1987,Linder2005}.
For HEED backgrounds,
\begin{equation}
\frac{d\ln H}{d\ln a}
=\frac{1}{2}\frac{d\ln Q^2}{d\ln a}
=-\frac{3}{2}\,\Omega_m(a)\;+\;\frac{1}{2}\,\frac{d c_e^2/d\ln a}{1-c_e^2}\,,
\label{eq:dlnH}
\end{equation}
so the “friction” term in Eq.~\eqref{eq:heed-growth-ode} is slightly enhanced during the HEED switch on ($dc_e^2/d\ln a>0$), yielding a mild suppression of growth relative to $\Lambda$CDM at $z\!\lesssim\!1$. Once $c_e^2(a)$ saturates ($d c_e^2/d\ln a\simeq 0$) the growth history closely tracks that of a smooth dark energy model with the same $Q(a)$. A useful approximation for intuition is the growth index form $f(a)\simeq \Omega_m(a)^\gamma$ with $\gamma\simeq 0.55$ in GR and small, $w(a)$ dependent corrections for smooth dark energy \cite{Linder2005}. The HEED behaves like such a smooth sector and therefore predicts percent level shifts in $f\sigma_8(z)$ concentrated at low redshift (as illustrated in~Fig.~\ref{fig:fs8}).

These features have clear observational implications. At $a=1$, Eq.~\eqref{eq:E2} gives $Q^{2}(1)=1/(1-c_{e0}^{2})$, so the inferred Hubble constant is boosted by $H_{0}^{\rm(eff)}/H_{0}=(1-c_{e0}^{2})^{-1/2}$, as illustrated in~Fig.~\ref{fig:Hz}. A modest $c_{e0}^{2}=0.06$ yields a $\sim 3.1\%$ increase, while $c_{e0}^{2}=0.10$ gives $\sim 5.3\%$. Because $g(a\ll a_{t})\ll 1$, recombination era distances and the sound horizon remain essentially unchanged, providing exactly the lever arm required by the early--late $H_{0}$ discrepancy. Low--redshift distance indicators (SN/BAO) can be matched with gentle adjustments driven by the amplitude of $c_{e0}^{2}$ and parameters $a_{t}$ and $k$ describing the activation history. A practical data strategy is to fit SN+BAO+CC backgrounds with a CMB distance prior to keep the sound
horizon $r_{d}$ standard, then test RSD $f\sigma_{8}(z)$~\cite{Kaiser1987,Hamilton1998},
enforcing the early time prior $c_e^{2}(a_{\star})\ll 1$ at recombination.

The construction developed above is restricted to homogeneous, spatially flat
FLRW cosmology and the apparent (Hubble) horizon. No fully covariant extension
to generic spacetimes is assumed here; the present aim is instead to formulate
and test the corresponding effective infrared cosmological dynamics in the
simplest setting relevant to the Hubble tension.

Finally, we note relationships and distinctions with neighboring ideas. Modular Hamiltonian first law analyses for ball regions show that free or conformal fields produce $\delta\rho\sim H^{4}$ corrections~\cite{Casini2011,BKS2016,Bhattacharya2013,Blanco2013,Herzog2013,Jacobson2013}
(see Appendix~\ref{sec:H4-scaling} for details), far too small to be cosmologically relevant; HEED avoids this suppression by tying the effect to IR horizon thermodynamics. Standard holographic dark energy simply posits $\rho\propto H^{2}/G$ with a constant coefficient; HEED explains that coefficient as a time dependent horizon entanglement shortfall $c_e^{2}(a)=2\,\delta(a)$ with a natural late time activation. The framework is local in time (apparent horizon) and thus avoids teleological issues
associated with future event horizons~\cite{Gibbons1977,Padmanabhan2010RPP,Padmanabhan2012,Wall2018}. In sum, HEED constitutes a thermodynamically
motivated, IR entanglement mechanism that is internally consistent (conservation, stability, horizon thermodynamics), naturally small at high redshift, and predictive for late time expansion and linear growth~\cite{Wall2018}.

\section{Observational consistency analysis}
\label{sec:observations}

In this section we confront the HEED activation parameter triplet with a combination of four low-$z$ probes, treated as independent Gaussian constraints. We use SN~Ia distance moduli $\mu(z)$ from Pantheon+~\cite{Scolnic2022PantheonPlusData,Brout2022PantheonPlusCosmo}; cosmic–chronometer (CC) determinations of $H(z)$ from differential age measurements of passively evolving early-type galaxies~\cite{zhang2014four,Simon:2005,M_Moresco_2012,Moresco_2016,MorescoMNRAS2015,Ratsimbazafy_2017,Daniel_Stern_2010};
compressed BAO measurements of the transverse comoving distance $d_M(z)/r_d$ and the radial comoving Hubble distance $d_H(z)/r_d$ from BOSS~DR12~\cite{1607.03155}, eBOSS~DR16~\cite{Alam2021eBOSS} and DESI~DR2~\cite{DESI2025DR2_BAO},
adopting a fixed sound horizon $r_d$~\cite{Planck2018}; and RSD measurements of the growth rate $f\sigma_8(z)$ from BOSS/eBOSS~\cite{1312.4854,1606.00439,1312.4889,1607.03155,1607.03150,1709.05173,1203.6565}, together with WiggleZ~\cite{1204.3674}, VIPERS~\cite{1303.2622,1610.08362,1611.07046,1612.05647,1612.05645,1708.00026}, supplemented by very low-$z$ ~\cite{1706.05130,1011.3114,1203.4814,1204.4725,1611.09862} and high-$z$~\cite{1511.08083,1801.03043,1801.02689,1801.02656} constraints.

For the CC sample we employ the community “classic’’ 31-point set, typically treated as uncorrelated, assembled from DA measurements (see Table~\ref{tab:CC_uncorr_31} in Appendix~\ref{data}); in this approach the redshift–age relation yields $dz/dt$, which is converted to the Hubble parameter via $H(z)=-(1+z)^{-1}\,dz/dt$. The BAO observables used here are compiled from Refs.~\cite{1607.03155,Alam2021eBOSS,DESI2025DR2_BAO}, with the full list of points provided in Table~\ref{tab:BAO_combined} of Appendix~\ref{data}.
Finally, Table~\ref{tab:fs8_tableI}  of Appendix~\ref{data} presents a consolidated set of 63 RSD measurements of the linear growth rate, reported as $f\sigma_8(z)$ with 1$\sigma$ uncertainties and spanning $z\simeq 0.001$–$1.944$, compiled in Ref.~\cite{RSDcompile63}. This compilation includes very low–$z$ peculiar velocity and Tully–Fisher constraints from 2MTF~\cite{1706.05130}, 2MRS~\cite{1011.3114,1203.4814}, and 6dFGS/6dFGS+SNe~\cite{1204.4725,1611.09862}; early RSD determinations from SDSS–LRG~\cite{JCAP.0910.004} (including the SDSS–LRG–200/60 re-analyses~\cite{1102.1014}) and DR7–LRG~\cite{1209.0210}; low–$z$ galaxy–survey results from SDSS–MGS~\cite{1409.3238}, SDSS–veloc~\cite{1503.05945}, DR13~\cite{1612.07809}, and GAMA~\cite{1309.5556}; the WiggleZ three–point set~\cite{1204.3674}; extensive BOSS measurements—LOWZ/CMASS and DR10–DR12 from Refs.~\cite{1312.4854,1606.00439,Alam2021eBOSS,1312.4889,1607.03155,1607.03150,1709.05173,1203.6565}; multiple VIPERS determinations (including v7 and PDR–2)~\cite{1303.2622,1610.08362,1611.07046,1612.05647,1612.05645,1708.00026}; additional SDSS DR7 peculiar–velocity work~\cite{1712.04163}; the FastSound high–$z$ constraint~\cite{1511.08083}; and the SDSS–IV QSO tomographic measurements at $z=0.978,\,1.23,\,1.526,\,1.944$~\cite{1801.03043,1801.02689,1801.02656}.

Distances follow the standard flat FLRW relations with $H(z)$ defined in Eq.~\eqref{eq:activation} and Eq.~\eqref{eq:E2}:
\begin{equation}
d_C(z)=\int_0^z\frac{c\,dz'}{H(z')},\qquad
d_M(z)=d_C(z),\qquad
d_L(z)=(1+z)d_M(z),\qquad
\mu(z)=5\log_{10}\!\big(D_L/{\rm Mpc}\big)+25.
\label{eq:distances}
\end{equation}
BAO observables are $d_M/r_d$ and $d_H/r_d$ with $d_H\equiv c/H$.

We analyze HEED in the \emph{additive embedding} at the level of the homogeneous
background, in which the entanglement sector contributes an algebraic
\(H^2\)-proportional term to the Friedmann equation expressed by Eq.~(\ref{eq:Ha0}).
Eliminating \(\rho_{\rm HEED}\) gives the closed background relation expressed by Eq.~(\ref{eq:E2}).
We normalize the activation so that \(c_e^2(a{=}1)=c_{e0}^2\), and parameterize
\begin{equation}
c_e^2(a)=c_{e0}^2\,g(a),
\end{equation}
with \(g(1)=1\) and a late-time switch-on controlled by \((a_t,k)\), as in
Eq.~(\ref{eq:activation}). In this embedding it is convenient to define the
\emph{effective} late-time Hubble scale,
\begin{equation}
H_0^{\rm eff}\equiv H(a{=}1)=\frac{H_0}{\sqrt{1-c_{e0}^2}},
\label{eq:H0eff_def}
\end{equation}
which is the quantity constrained by local measurements.
Linear growth assumes a smooth, non-clustering HEED sector in general relativity (GR) so that the growth factor $D(a)$ satisfies Eq.~(\ref{eq:heed-growth-ode}) integrated from matter domination with $D(a{=}1)=1$. The RSD prediction is
\begin{equation}
f\sigma_8(z)=\left.\frac{d\ln D}{d\ln a}\right|_{a=1/(1+z)} \sigma_{8,0}\,D\!\left(a=\frac{1}{1+z}\right),
\label{eq:rsd1}
\end{equation}
with $\sigma_{8,0}$ held fixed as a Gaussian prior.

\subsection{Bayesian formulation}
\label{sec:bayes}
The baseline parameter vector in use is
\begin{equation}
\boldsymbol{\theta}=\{H_0,\Omega_m,c_{e0}^2,a_t,k,\sigma_8\},
\end{equation}
where \(H_0\) is the underlying Hubble constant entering
Eq.~(\ref{eq:E2}), while \(H_0^{\rm eff}\) is given by
Eq.~(\ref{eq:H0eff_def}). The data vector
\(
\mathbf{d}=\{\boldsymbol{\mu}^{\,\rm obs},\,\mathbf{b}^{\,\rm obs},\,\mathbf{h}^{\,\rm obs},\,\mathbf{g}^{\,\rm obs}\}
\)
collects SN\,Ia distance moduli \(\boldsymbol{\mu}^{\,\rm obs}\),
BAO points \(\mathbf{b}^{\,\rm obs}\equiv\{D_M/r_d,\,D_H/r_d\}\),
CC measurements \(\mathbf{h}^{\,\rm obs}\equiv\{H(z)\}\),
and RSD growth measurements \(\mathbf{g}^{\,\rm obs}\equiv\{f\sigma_8(z)\}\).
For a given \(\boldsymbol{\theta}\), the corresponding predictions
\(\boldsymbol{\mu}^{\,\rm th}(\boldsymbol{\theta})\),
\(\mathbf{b}^{\,\rm th}(\boldsymbol{\theta})\),
\(\mathbf{h}^{\,\rm th}(\boldsymbol{\theta})\),
and \(\mathbf{g}^{\,\rm th}(\boldsymbol{\theta})\)
are computed from \(H(z;\boldsymbol{\theta})\) via
Eqs.~(\ref{eq:distances}), (\ref{eq:heed-growth-ode}), and (\ref{eq:rsd1}).

We define the posterior as
\begin{equation}
\label{Poster1b}
\mathcal{P}(\boldsymbol{\theta}\mid \mathbf{d})
\ \propto\
\mathcal{L}(\mathbf{d}\mid \boldsymbol{\theta})\ \Pi(\boldsymbol{\theta}),
\qquad
\ln \mathcal{L} = -\tfrac{1}{2}\,\chi^2_{\rm tot}(\boldsymbol{\theta}),
\end{equation}
with the total quadratic statistic
\begin{equation}
\label{chiTot1b}
\chi^2_{\rm tot}(\boldsymbol{\theta})
=\chi^2_{\rm SN}
+\chi^2_{\rm BAO}
+\chi^2_{\rm CC}
+\chi^2_{\rm RSD}
+\chi^2_{\sigma_8}
+\chi^2_{\rm prior},
\end{equation}
assuming Gaussian errors and (for this baseline analysis) independence between
data blocks.

\begin{figure}[htbp]
  \centering
  \includegraphics[width=0.8\linewidth]{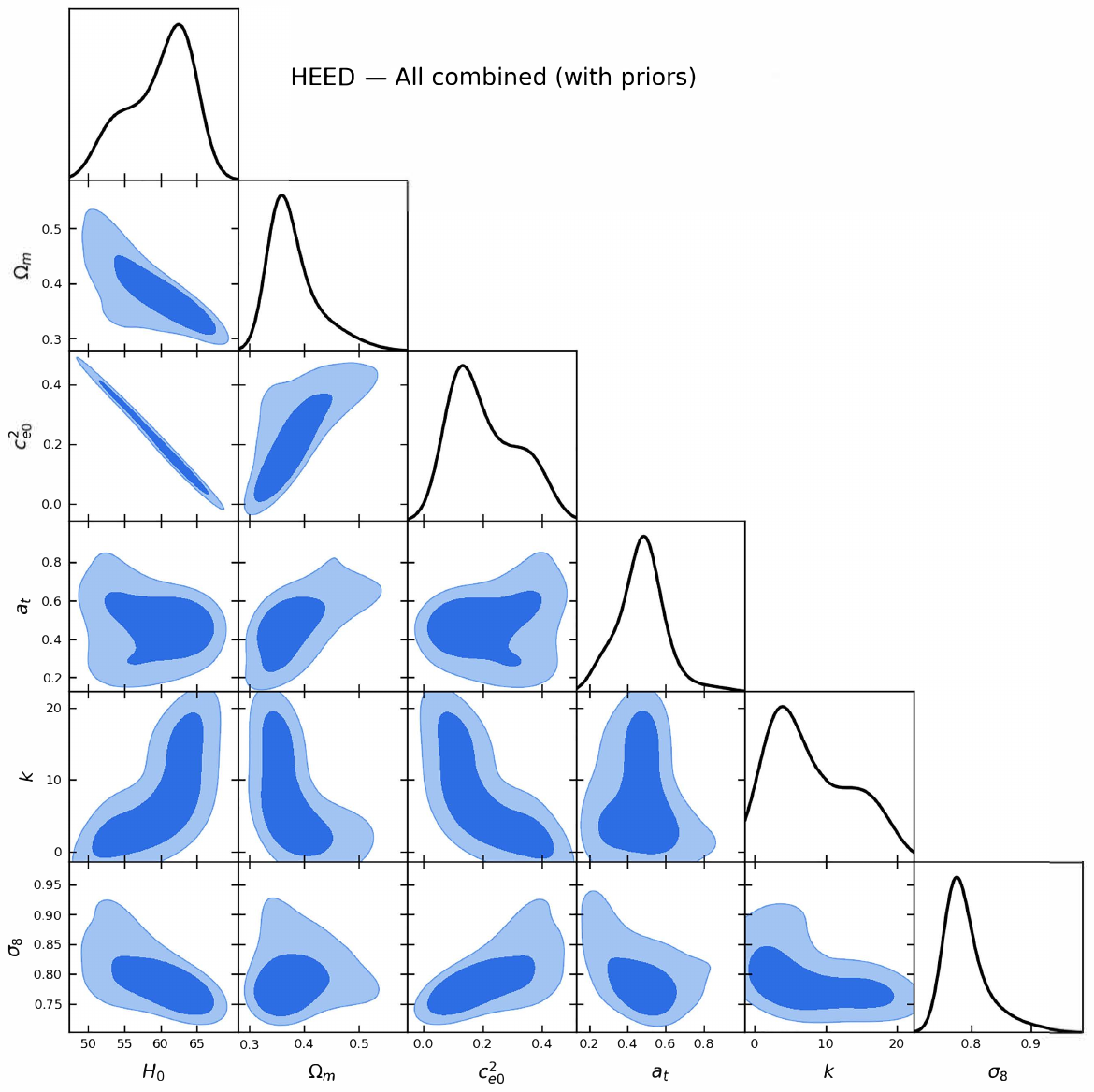}
  \caption{Triangle plot for the HEED parameterization showing the marginalized
posterior distributions of $\{H_0,\Omega_m,c_{e0}^2,a_t,k,\sigma_8\}$ obtained
from the joint SN+BAO+CC+RSD likelihood. Shaded contours indicate the 68\% and 95\% credible
regions, while the diagonal panels show the corresponding one--dimensional
marginals. In the anchored analysis, the external Gaussian prior is applied to the
effective late--time Hubble scale $H_0^{\rm eff}=H_0/\sqrt{1-c_{e0}^2}$ rather
than directly to $H_0$, which induces the characteristic degeneracy direction
in the $(H_0,c_{e0}^2)$ plane.}

\label{fig:triangle1}
\end{figure}

\begin{figure}[t]
\centering
\begin{minipage}{0.48\linewidth}
    \includegraphics[width=\linewidth]{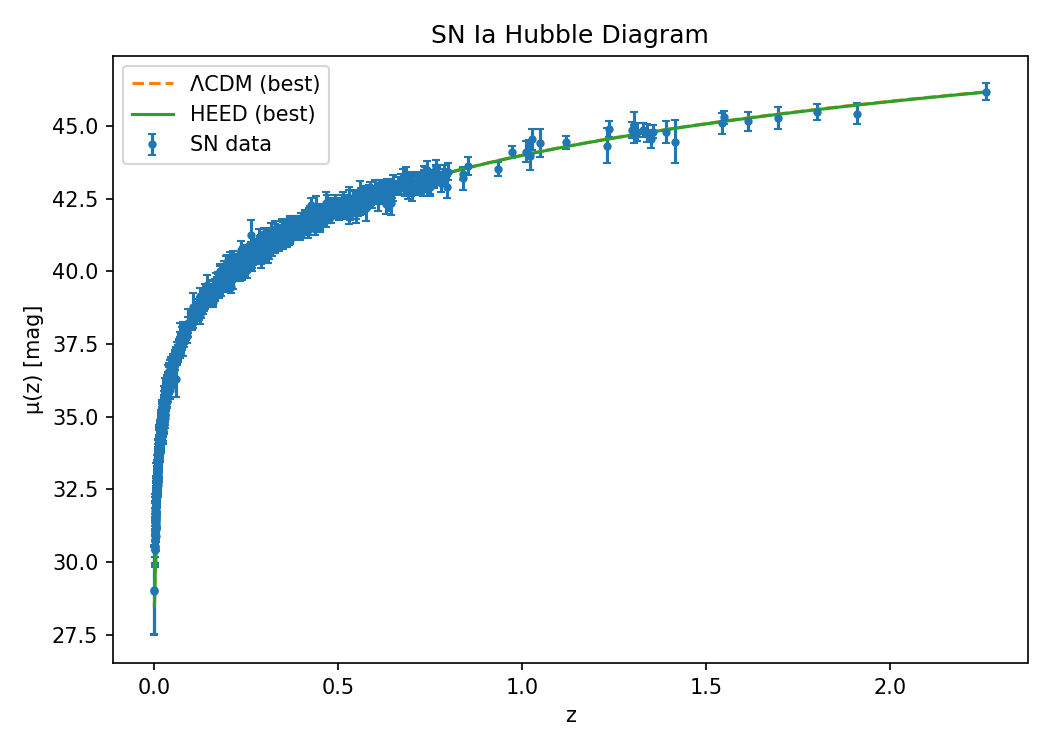}
    \caption{Pantheon+ SN~Ia distance moduli $\mu(z)$ compared to the best--fit
HEED (solid) and $\Lambda$CDM (dashed) predictions. The fits are shown after
profiling over the nuisance offset $\Delta M$ (absolute-magnitude/zero-point)
as described in the text.}
    \label{fig:SN}
\end{minipage}
\hfill
\begin{minipage}{0.48\linewidth}
    \includegraphics[width=\linewidth]{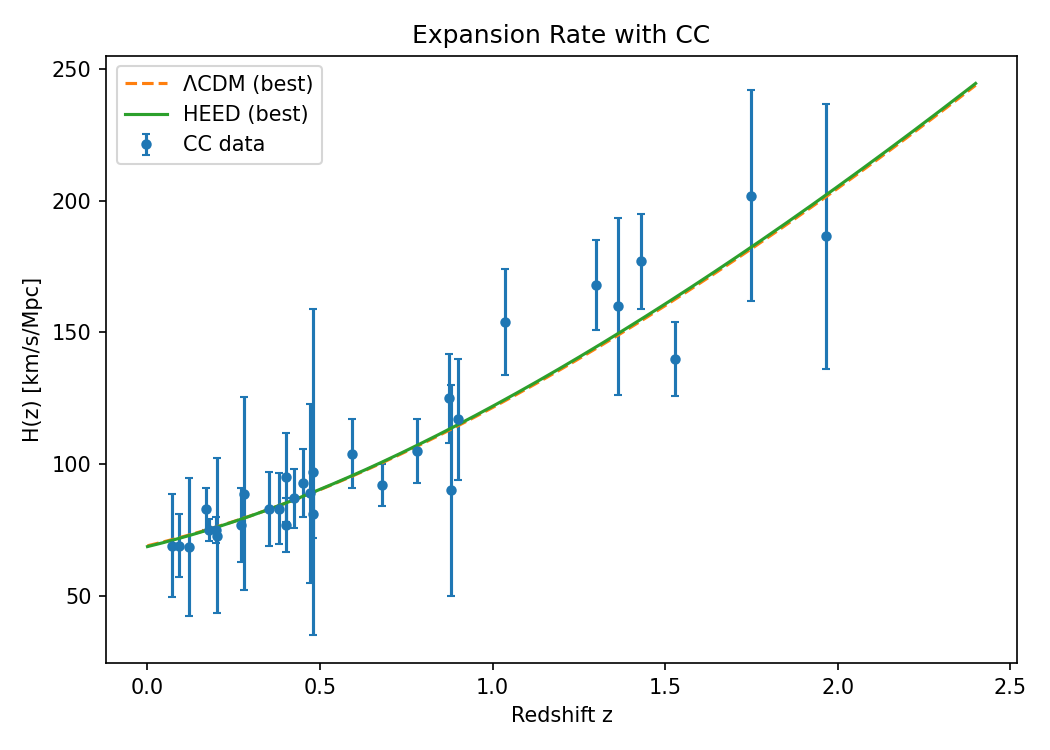}
    \caption{CC measurements of $H(z)$ with $1\sigma$
uncertainties compared to the best--fit HEED (solid) and $\Lambda$CDM (dashed)
predictions.}
    \label{fig:CC}
\end{minipage}
\end{figure}

\begin{figure}[t]
\centering
\begin{minipage}{0.48\linewidth}
    \includegraphics[width=\linewidth]{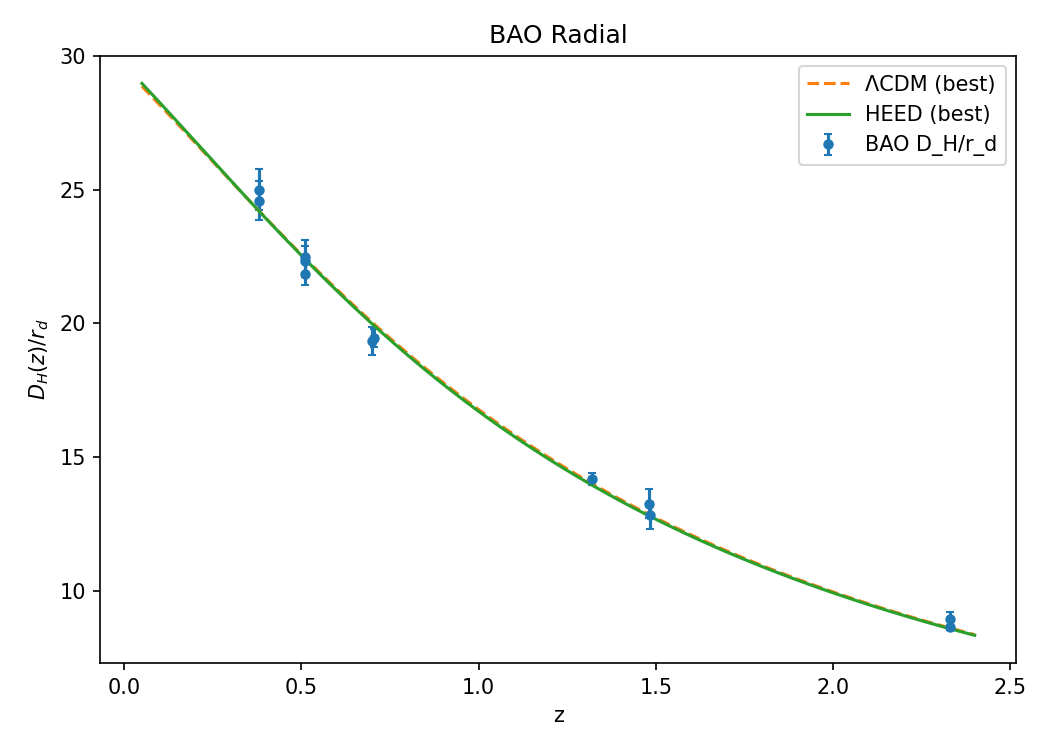}
    \caption{BAO measurements of the radial (line-of-sight) comoving distance,
$D_H(z)/r_d$ with $D_H\equiv c/H(z)$, compared to the best--fit HEED (solid) and
$\Lambda$CDM (dashed) predictions.}
    \label{fig:BAOr}
\end{minipage}
\hfill
\begin{minipage}{0.48\linewidth}
    \includegraphics[width=\linewidth]{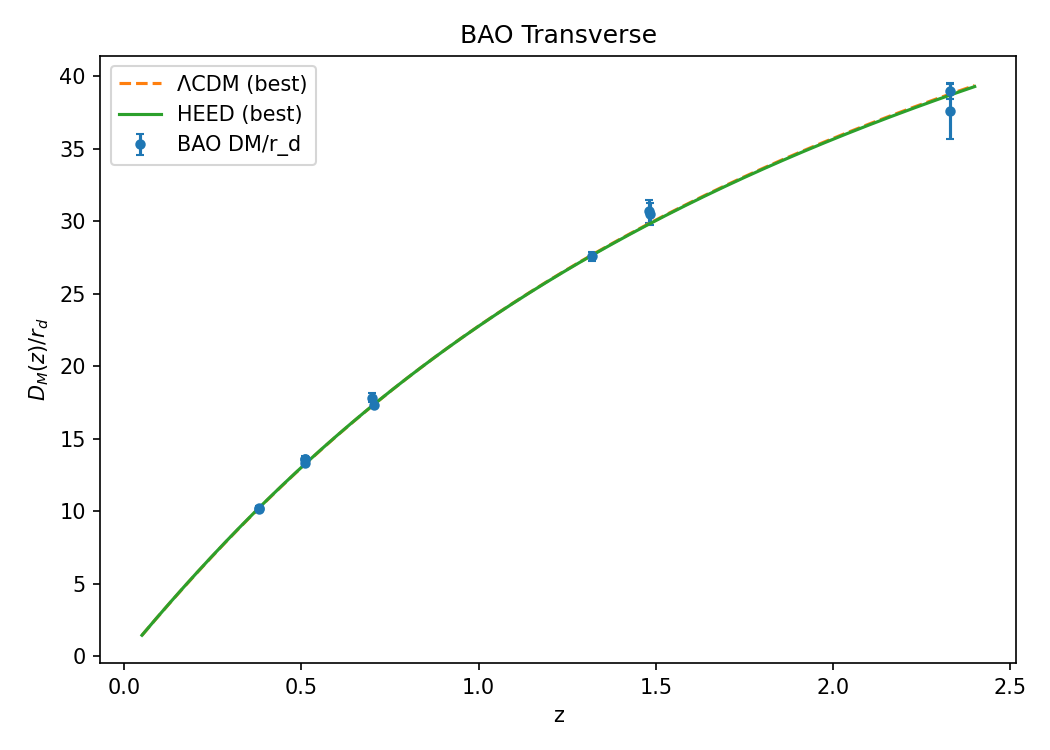}
    \caption{BAO measurements of the transverse comoving distance, $D_M(z)/r_d$,
compared to the best--fit HEED (solid) and $\Lambda$CDM (dashed) predictions.}
    \label{fig:BAOt}
\end{minipage}
\end{figure}

\begin{figure}[htbp]
  \centering
  \includegraphics[width=0.8\linewidth]{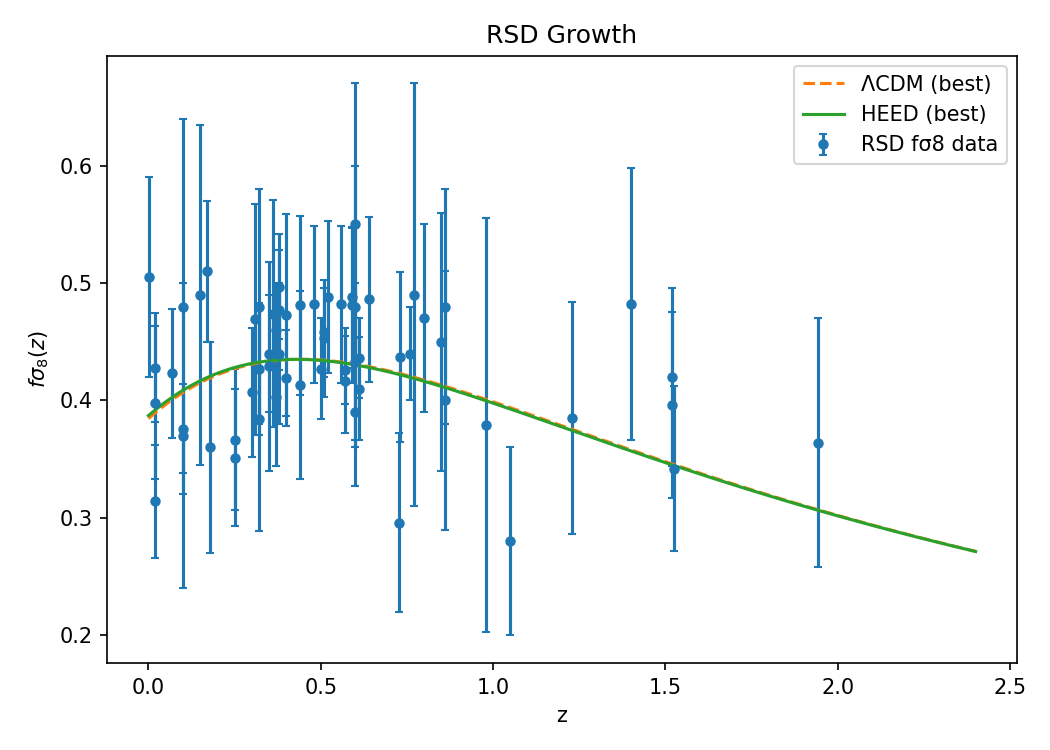}
  \caption{RSD growth constraints in the HEED framework.
Measurements of $f\sigma_8(z)$ with $1\sigma$ uncertainties compared to the
best--fit predictions for HEED (additive embedding) and $\Lambda$CDM.
Model curves are obtained by solving the linear growth equation
Eq.~(\ref{eq:heed-growth-ode}) assuming a smooth (non--clustering) HEED sector,
normalizing $D(a{=}1)=1$, and computing
$f\sigma_8(z)=\big(d\ln D/d\ln a\big)\,\sigma_{8,0}\,D(a)$ at $a=(1+z)^{-1}$.}
\label{fig:RSD}
\end{figure}

The SN term uses Hubble diagram residuals and profiles out a single additive
offset \(\Delta M\), encoding the absolute magnitude (distance zero-point)
degeneracy with the Hubble scale. With diagonal errors the raw statistic is
\begin{equation}
\chi^2_{\rm SN}(\boldsymbol{\theta},\Delta M)
=\sum_i \frac{\big[\mu_i-\mu_{\rm th}(z_i;\boldsymbol{\theta})-\Delta M\big]^2}
{\sigma_{\mu,i}^2}\,,
\label{eq:chisq-sn-raw-bayes}
\end{equation}
where \(\mu_{\rm th}(z)\) is evaluated from \(H(z)\).
Minimization of Eq.~\eqref{eq:chisq-sn-raw-bayes} with respect to \(\Delta M\)
yields the analytic profiler
\begin{equation}
\Delta M^\star(\boldsymbol{\theta})
=\frac{\displaystyle\sum_i \big[\mu_i-\mu_{\rm th}(z_i;\boldsymbol{\theta})\big]/\sigma_{\mu,i}^2}
{\displaystyle\sum_i 1/\sigma_{\mu,i}^2}\,,
\qquad
\chi^2_{\rm SN}(\boldsymbol{\theta})\equiv
\chi^2_{\rm SN}\big(\boldsymbol{\theta},\Delta M^\star(\boldsymbol{\theta})\big),
\label{eq:deltaM-profile-bayes}
\end{equation}
which is equivalent to analytically marginalizing over a constant offset with a
flat prior. Operationally, we profile separately for each cosmological model
(HEED and \(\Lambda\)CDM), so each curve is shifted by its own
\(\Delta M^\star\), consistent with standard practice in compressed SN
likelihood applications \cite{Scolnic2022PantheonPlusData,Brout2022PantheonPlusCosmo}.
Consequently, SN data alone do not fix \(H_0\); the absolute scale is set only
when combined with an absolute ruler such as CC or BAO (through \(r_d\)), or by
including an explicit external anchor.

The BAO and CC blocks adopt diagonal Gaussian forms,
\begin{align}
\chi^2_{\rm BAO} &=
\sum_j \left[
\frac{\big(d_M(z_j)/r_d-(d_M/r_d)_{\rm obs}\big)^2}{\sigma_{d_M/r_d,j}^2}
+\frac{\big(d_H(z_j)/r_d-(d_H/r_d)_{\rm obs}\big)^2}{\sigma_{d_H/r_d,j}^2}
\right],\label{eq:chi2_bao_bayes}\\
\chi^2_{\rm CC} &=
\sum_\ell \frac{\big(H(z_\ell)-H_{\rm obs}\big)^2}{\sigma_{H,\ell}^2},
\label{eq:chi2_cc_bayes}
\end{align}
which are the diagonal-covariance limits of the standard Gaussian likelihood
\begin{equation}
\chi^2
=
\Delta\mathbf{d}^{\,T}\mathbf{C}^{-1}\Delta\mathbf{d},
\qquad
\Delta\mathbf{d}
\equiv
\mathbf{d}^{\rm obs}
-
\mathbf{d}^{\rm th}(\boldsymbol{\theta}),
\label{eq:full_covariance_chi2_insert}
\end{equation}
where \(\mathbf{C}\) is the covariance matrix of the corresponding data vector.
Equations~\eqref{eq:chi2_bao_bayes} and~\eqref{eq:chi2_cc_bayes} therefore
amount to taking \(\mathbf{C}\) to be diagonal, with entries given by the
reported variances.

The corresponding grows log-likelihood is taken as
\begin{equation}
\chi^2_{\rm RSD}
=\sum_m \frac{\big[f\sigma_8(z_m;\boldsymbol{\theta})-(f\sigma_8)_{\rm obs}\big]^2}
{\sigma_{f\sigma_8,m}^2}\,,
\label{eq:chi2_rsd_bayes}
\end{equation}
also with diagonal errors in this baseline implementation.
In practice, the RSD block provides the primary handle on late--time structure
growth in our low--$z$ data combination, since the predicted $f\sigma_8(z)$ is
obtained by solving the linear growth equation with coefficients determined by
the background expansion $H(z)$. Within the assumptions of GR growth and a
smooth (non--clustering) HEED sector, this term therefore links the background
parameters controlling $H(z)$ to the growth history encoded in $D(a)$ and
$f\equiv d\ln D/d\ln a$.

To regularize the growth sector, we include a weak Gaussian prior on
\(\sigma_8\),
\begin{equation}
\chi^2_{\sigma_8}
=\left(\frac{\sigma_8-\sigma_{8,\star}}{\sigma_{\sigma_8}}\right)^2,
\label{eq:chi2_sigma8_bayes}
\end{equation}
which may be chosen broad enough to avoid importing CMB-dominated information,
yet prevents unphysical excursions when the RSD block alone is insufficient to
pin down \(\sigma_8\) in combination with \((\Omega_m,c_{e0}^2,a_t,k)\).

We include three priors/guards, assembled as
\begin{equation}
\chi^2_{\rm prior}
=\chi^2_{\rm early}
+\chi^2_{\rm wall}
+\chi^2_{H_0^{\rm eff}}.
\label{eq:chi2_prior_sum_bayes}
\end{equation}
The early-time prior enforces negligible HEED at recombination,
\begin{equation}
\chi^2_{\rm early}=
\begin{cases}
0, & c_e^2(a_\star)\le \varepsilon,\\[4pt]
\displaystyle \left(\frac{c_e^2(a_\star)-\varepsilon}{\varepsilon}\right)^2 W,
& c_e^2(a_\star)>\varepsilon,
\end{cases}
\qquad
a_\star=(1+1100)^{-1},\ \varepsilon=10^{-3},\ W=10^4,
\label{eq:chi2_early_bayes}
\end{equation}
and the wall prior penalizes approaching \(c_e^2\to1\) (where the effective
Planck mass \(M_*^2\propto 1-c_e^2\) would vanish) over the redshift range of the
data,
\begin{equation}
\chi^2_{\rm wall}=
\begin{cases}
0, & \displaystyle \max_{a\in[a_{\rm min},1]} c_e^2(a)\le c_{\rm wall}^2,\\[6pt]
\displaystyle W_{\rm wall}\,\big(\max c_e^2-c_{\rm wall}^2\big)^2, & \text{otherwise},
\end{cases}
\label{eq:chi2_wall_bayes}
\end{equation}
with \(a_{\rm min}=(1+z_{\max})^{-1}\) and \(z_{\max}\) the largest redshift used
in the likelihood.

Finally, in anchored analysis we impose an external Gaussian prior on the
\emph{effective} Hubble scale \(H_0^{\rm eff}\), not on \(H_0\),
\begin{equation}
\chi^2_{H_0^{\rm eff}}
=\left(\frac{H_0^{\rm eff}-H_{0,\star}^{\rm eff}}{\sigma_{H_0^{\rm eff}}}\right)^2,
\qquad
H_0^{\rm eff}=\frac{H_0}{\sqrt{1-c_{e0}^2}}.
\label{eq:chi2_H0eff_bayes}
\end{equation}
This prior is not used to infer the Hubble scale from low-$z$ probes alone.
Rather, it enables a consistency test: we ask whether HEED can accommodate a
locally inferred high late-time expansion rate while remaining compatible with
SN, BAO, CC, and RSD constraints that jointly probe distances and growth.

In the anchored runs we impose an external Gaussian prior on the effective
late--time Hubble scale \(H_0^{\rm eff}\), not directly on the sampled parameter
\(H_0\). In the additive HEED embedding,
\begin{equation}
H_0^{\rm eff}=\frac{H_0}{\sqrt{1-c_{e0}^2}} .
\end{equation}
Thus the analysis should be interpreted as a consistency test: it asks whether
HEED can accommodate a locally inferred high late--time expansion rate while
remaining compatible with SN, BAO, CC, and RSD constraints. It does not by
itself constitute a free low--redshift derivation of a high \(H_0\).

We sample the posterior \(\mathcal{P}(\boldsymbol{\theta}\mid\mathbf{d})\) using
the affine-invariant ensemble sampler \texttt{emcee}~\cite{emcee0}, and compute marginalized
constraints and triangle plots using \texttt{GetDist}~\cite{GetDist0}.

For a simple model-comparison diagnostic, we quote the Akaike and Bayesian
information criteria \cite{Akaike1974,Schwarz1978,Liddle2004,Liddle2007},
\begin{equation}
{\rm AIC}
=
\chi^2_{\rm min}+2N_{\rm par},
\qquad
{\rm BIC}
=
\chi^2_{\rm min}+N_{\rm par}\ln N_{\rm data},
\label{eq:AIC_BIC_insert}
\end{equation}
where \(N_{\rm par}\) is the number of fitted parameters and \(N_{\rm data}\) is
the number of data points entering the simplified likelihood. We report
\(\Delta{\rm AIC}\) and \(\Delta{\rm BIC}\) relative to the corresponding
\(\Lambda\)CDM fit. Since the likelihood used here neglects some covariance
matrices, these values should be interpreted as indicative rather than as a
final Bayesian model selection result. The numerical results of the consistency analysis are collected in
Tables~\ref{tab:HEED_constraints} and~\ref{tab:model_comparison}.
Table~\ref{tab:HEED_constraints} reports the adopted priors, marginalized
constraints and best-fit values for the HEED parameters, while
Table~\ref{tab:model_comparison} gives the data set by data set
\(\chi^2\) contributions and the corresponding \(\Delta\chi^2\), AIC and BIC
comparison with \(\Lambda\)CDM. These model-comparison indicators are computed
within the simplified diagonal likelihood and should therefore be regarded as
indicative rather than definitive.

\begin{table*}[t]
\caption{
Summary of priors, marginalized constraints, and best-fit values for the HEED
analysis. The quoted intervals correspond to the marginalized 68\% credible ranges.
Approximate marginalized constraints read from the one-dimensional posterior
marginals in Fig.~\ref{fig:triangle1}.
}
\label{tab:HEED_constraints}
\begin{ruledtabular}
\begin{tabular}{lcc}
Parameter & Prior & Marginalized constraint \\
\hline

$H_0^{\rm eff}\,[{\rm km\,s^{-1}\,Mpc^{-1}}]$
& external Gaussian anchor
& $\simeq 73.0\pm 1.0$ \\

$\Omega_m$
& --
& $0.365^{+0.057}_{-0.026}$ \\

$c_{e0}^2$
& $0\leq c_{e0}^2<1$
& $0.16^{+0.19}_{-0.07}$ \\

$a_t$
& late-time activation prior
& $0.47^{+0.10}_{-0.13}$ \\

$k$
& $k>0$
& $5.8^{+9.6}_{-3.8}$ \\

$\sigma_8$
& weak Gaussian prior
& $0.779^{+0.045}_{-0.021}$ \\
\end{tabular}
\end{ruledtabular}
\end{table*}

\begin{table*}[t]
\caption{
Goodness of fit and information criterion comparison between HEED and
$\Lambda$CDM. The values are computed with the simplified diagonal likelihood
used in this work and should therefore be interpreted as indicative.
}
\label{tab:model_comparison}
\begin{ruledtabular}
\begin{tabular}{lccccccc}
Model
& $\chi^2_{\rm SN}$
& $\chi^2_{\rm BAO}$
& $\chi^2_{\rm CC}$
& $\chi^2_{\rm RSD}$
& $\chi^2_{\rm tot}$
& AIC
& BIC \\
\hline
$\Lambda$CDM
& $750$
& $21.6$
& $14.7$
& $32.5$
& $819$
& $825$
& $842$ \\

HEED
& $745$
& $19.2$
& $15.1$
& $32.1$
& $811$
& $823$
& $856$ \\

\hline
HEED $-\Lambda$CDM
& $-5$
& $-2.4$
& $0.4$
& $-0.4$
& $\Delta\chi^2=-8$
& $\Delta{\rm AIC}=-2$
& $\Delta{\rm BIC}=14$ \\
\end{tabular}
\end{ruledtabular}
\end{table*}

The HEED model, as indicated in Table~\ref{tab:model_comparison}, reduces the total data chi-square
by $\Delta \chi^2_{\rm tot} \simeq -8$ .
This improvement is driven mainly by the supernova and BAO contributions,
$\Delta \chi^2_{\rm SN} \simeq -5.0, ~ \Delta \chi^2_{\rm BAO} \simeq -2.4$,
whereas the changes in the CC and RSD sectors are small, $\Delta \chi^2_{\rm CC} \simeq +0.4,
~ \Delta \chi^2_{\rm RSD} \simeq -0.4$ . Thus the raw goodness of fit is mildly improved relative to \(\Lambda\)CDM.
However, HEED contains three additional free parameters relative to
\(\Lambda\)CDM. The corresponding AIC difference is
\[
\Delta{\rm AIC}
\equiv
{\rm AIC}_{\rm HEED}
-
{\rm AIC}_{\Lambda{\rm CDM}}
=
\Delta\chi^2_{\rm tot}
+
2\Delta N_{\rm par}
\simeq
-8+2\times3
\simeq -2 .
\]
A negative value of \(\Delta{\rm AIC}\) indicates that the improvement in
\(\chi^2\) only mildly compensates for the additional HEED parameters. In
contrast, the BIC difference is
\[
\Delta{\rm BIC}
\equiv
{\rm BIC}_{\rm HEED}
-
{\rm BIC}_{\Lambda{\rm CDM}}
=
\Delta\chi^2_{\rm tot}
+
\Delta N_{\rm par}\ln N_{\rm data}
\simeq
-8+3\ln(1819)
\simeq 14 .
\]
The positive value of \(\Delta{\rm BIC}\) favors \(\Lambda\)CDM, reflecting the
stronger BIC penalty for additional parameters when the number of data points is
large. We therefore interpret the fit as an indicative improvement in raw
goodness of fit, but not as a robust model selection preference for HEED.


The likelihood adopted in this exploratory analysis treats the compressed SN,
BAO, CC, and RSD data points as independent Gaussian measurements. Equivalently,
the full covariance expression in Eq.~\eqref{eq:full_covariance_chi2_insert} is
approximated by retaining only the diagonal variances. Neglecting off-diagonal
covariances can misestimate both the constraining power of the data and the
relative weights of the different data blocks, especially for Pantheon+,
correlated BOSS/eBOSS/DESI BAO products, and RSD compilations assembled from
partially overlapping galaxy samples.

As a first diagnostic, we repeated the fit including the cosmic-chronometer
covariance induced by stellar-population-synthesis systematics, following
Ref.~\cite{Moresco:2020fbm}. The resulting change in the RSD-related fit quality
is at the level of approximately \(2\%\). This suggests that the qualitative
HEED trends are stable against this particular covariance source. Nevertheless,
modern large-scale-structure analyses show that covariance modelling can affect
projected uncertainties at the few percent level. For example, the DESI 2024
covariance-validation study of Ref.~\cite{ForeroSanchez:2024DESICovariance}
finds that analytical and mock-based covariance estimates agree well for BAO
distance-scale errors, while full-shape analyses require a more careful
covariance treatment.

Because Pantheon+, BAO, and RSD data products contain non-trivial internal and
cross-survey correlations, we conservatively expect a full covariance treatment
to affect selected \(\chi^2\) contributions and marginalized uncertainties at
the few-percent to \(\mathcal{O}(10\%)\) level. The present diagonal treatment
should therefore be regarded as a first consistency test rather than as a
definitive quantitative model-selection analysis. A full covariance treatment
will be required before claiming a statistical preference for HEED over
\(\Lambda\)CDM.

\subsection{Posterior constraints and internal consistency of HEED}
\label{sec:triangle}

Figure~\ref{fig:triangle1} presents the marginalized one-- and two--dimensional
posterior distributions for the HEED parameter vector
\(\{H_0,\Omega_m,c_{e0}^2,a_t,k,\sigma_8\}\), obtained from the joint
SN+BAO+CC+RSD likelihood with the priors and profiling procedure specified in
Sec.~\ref{sec:bayes}. The shaded contours enclose the 68\% and 95\% credible
regions, while the diagonal panels show the corresponding one--dimensional
marginals. Complementary to these parameter constraints, the data space fits
shown in Figs.~\ref{fig:SN}, \ref{fig:CC}, \ref{fig:BAOr}, \ref{fig:BAOt}, \ref{fig:RSD}
demonstrate that the best fit
HEED and $\Lambda$CDM predictions are nearly indistinguishable across the
low redshift observables considered here (Pantheon+ $\mu(z)$, CC $H(z)$, BAO
$D_M/r_d$ and $D_H/r_d$, and RSD $f\sigma_8$), with no systematic preference
visible at the level of current uncertainties. We therefore interpret the
posterior in Fig.~\ref{fig:triangle1} primarily as a consistency statement:
within the adopted additive HEED embedding and priors, a late time
entanglement driven modification of the expansion history can be introduced
without spoiling established low--$z$ distance and growth constraints.

A number of qualitative and quantitative features demonstrate that the inferred constraints are internally consistent with the physical assumptions underlying HEED.

\subsubsection{Hubble constant}
As pointed in Sec.~\ref{sec:bayes}, in the present analysis, the posterior is conditioned on an external Gaussian prior (``anchor'') on the effective late--time Hubble scale centered at \(H_{0,\star}^{\rm eff}=73~{\rm km\,s^{-1}\,Mpc^{-1}}\) as inferred from~\cite{Riess2022}.
Accordingly, the chains do not \emph{infer} an upward shift of the Hubble scale relative to CMB-only analyses; rather, they test whether the HEED background and growth sector can remain compatible with SN+BAO+CC+RSD while accommodating a high locally inferred expansion rate.
Because in the additive HEED embedding Eq.~(\ref{eq:H0eff_def}),
the anchor does not force the sampled
parameter $H_0$ itself to be large. Rather, the posterior realizes
$H_0^{\rm eff}\simeq 73~{\rm km\,s^{-1}\,Mpc^{-1}}$ with moderate values of $H_0$
provided $c_{e0}^2>0$, leading to a pronounced anti--correlation between
$H_0$ and $c_{e0}^2$ in the marginalized posterior shown in
Fig.~\ref{fig:triangle1}.

\subsubsection{Matter density}
The matter density parameter \(\Omega_m\) is constrained at the level typical of low-redshift analyses, with a clear anti-correlation with \(H_0\). This degeneracy is expected: an increase in the late-time expansion rate induced by HEED must be compensated by a corresponding adjustment in \(\Omega_m\) in order to preserve the BAO distance measurements.
The inferred values of the matter density parameter remain within the range
allowed by late time large scale structure (LSS) probes, including BAO geometry,
RSD measurements and galaxy clustering with weak lensing combinations. In particular, the moderate
increase in $\Omega_m$ relative to $\Lambda$CDM is compensated by the
HEED--induced suppression of late--time growth, so that the predicted
$f\sigma_8(z)$ remains consistent with current observational constraints.

\subsubsection{Present-day HEED amplitude.}
A central result of the analysis is that the posterior for the HEED amplitude \(c_{e0}^2\) peaks decisively away from zero. Pure \(\Lambda\)CDM, corresponding to \(c_{e0}^2=0\), lies in the tail of the posterior. At the same time, the data strongly disfavor values approaching the physical wall \(c_e^2\to 1\), ensuring a positive effective Planck mass \(M_*^2 \propto 1-c_e^2\) and the absence of instabilities. This provides direct phenomenological support for a nonvanishing late-time horizon entanglement deficit.

\subsubsection{Activation epoch and sharpness}
The activation scale factor \(a_t\) is constrained to lie in the range \(a_t\sim0.4\!-\!0.6\), corresponding to a redshift \(z_t\sim0.7\!-\!1.5\). This places the onset of HEED squarely in the late universe, well after recombination and the baryon-drag epoch. As a result, early universe observables such as the sound horizon remain unaffected, while the late-time expansion history is modified in precisely the redshift range probed by SN, BAO, and CC data.

The shape parameter \(k\), which controls the sharpness of the HEED switch-on, is only weakly constrained by current data and exhibits broad, possibly multi-modal posteriors. This behavior is expected: existing low-redshift probes are primarily sensitive to the integrated effect of the late-time expansion rate rather than to the detailed functional form of the transition. Consequently, \(k\) effectively plays the role of a nuisance parameter at present precision.

The weak constraints on the activation parameters $(a_t,k)$ have a simple
origin. The data used here are mostly sensitive to the integrated impact of the
HEED sector on distances and growth, rather than to the detailed time profile of
the transition. Schematically, the activation history enters through
$c_e^2(a)=c_{e0}^2\,g(a;a_t,k)$,
but SN and BAO observables depend mainly on integrals of the form
$D(z)\sim \int_0^z \frac{dz'}{H(z')}$,
while RSD measurements constrain combinations such as $f\sigma_8(z)$ that also
respond to the cumulative expansion history. Consequently, different pairs
$(a_t,k)$ can generate nearly the same integrated distance and growth response
once $c_{e0}^2$, $H_0$, and $\Omega_m$ are adjusted. The parameter $a_t$ is
therefore constrained mainly by the requirement that HEED remain inactive before
recombination and the baryon drag epoch, while the sharpness parameter $k$
remains only weakly determined by present low redshift data. More direct
constraints on the transition profile would require observables with finer
redshift resolution or additional probes of late time potentials, such as weak
lensing, CMB lensing, or ISW cross-correlations.

\subsubsection{Growth amplitude}
The posterior for \(\sigma_8\) is mildly shifted toward lower values compared to the
Planck \(\Lambda\)CDM determination~\cite{Planck2018}.
This reflects the fact that the enhanced late--time expansion induced by HEED leads to a modest suppression
of structure growth at \(z\lesssim 1\).
The resulting values of \(\sigma_8\) and the correlated predictions for \(f\sigma_8(z)\) remain compatible with RSD measurements,
while potentially easing the weak-lensing clustering-amplitude (\(S_8\)) tension, i.e.\ the preference of several stage-III cosmic-shear analyses for
\(S_8\equiv \sigma_8\sqrt{\Omega_m/0.3}\) values lower than the CMB-inferred Planck \(\Lambda\)CDM prediction
(e.g.\ KiDS-1000 and DES~Y3)~\cite{Asgari2021KiDS1000,Heymans2021KiDS1000,Secco2022DESY3}.

\subsubsection{Correlation structure}
The two--dimensional marginalized posteriors exhibit a pattern of correlations
that can be understood directly from the structure of the HEED background and
from the combination of low--redshift probes employed.
Because the analysis is conditioned on an external prior on the effective
late--time Hubble scale $H_0^{\rm eff}$, the sampled parameters $H_0$ and
$c_{e0}^2$ are strongly anti--correlated. This follows from the additive HEED
relation Eq.~(\ref{eq:H0eff_def}): for a fixed $H_0^{\rm eff}$, a larger
entanglement deficit requires a smaller underlying $H_0$. The corresponding
elongated contours in the $(H_0,c_{e0}^2)$ plane therefore reflect a purely
kinematic degeneracy rather than a physical instability.

A positive correlation is observed between $c_{e0}^2$ and $\Omega_m$. In the
anchored analysis, increasing $c_{e0}^2$ lowers the sampled value of $H_0$,
which tends to increase comoving distances at fixed dimensionless expansion
rate. To preserve the BAO and CC constraints on distances and
$H(z)$, the fit compensates by increasing $\Omega_m$, which raises $Q(z)$ at
intermediate redshifts and reduces the integrated distances. This correlation
therefore encodes the geometric balance required to maintain consistency with
BAO measurements in the presence of a late--time entanglement contribution.

The correlation between $c_{e0}^2$ and $\sigma_8$ reflects the expansion--growth
interplay probed by RSDs. A larger entanglement deficit
enhances the late--time expansion rate, which mildly suppresses the growth of
linear perturbations at $z\lesssim1$. This suppression can be partially offset
by a correlated adjustment of the present--day fluctuation amplitude
$\sigma_8$, ensuring that the predicted $f\sigma_8(z)$ remains compatible with
RSD data. As a result, the joint posterior occupies a region in parameter space
where background expansion and structure growth are simultaneously consistent.

Finally, the activation parameters $(a_t,k)$ exhibit broad degeneracies with
$c_{e0}^2$ and with each other. Current low--redshift data primarily constrain
the integrated effect of the HEED contribution on the expansion history rather
than the detailed time profile of its onset. Consequently, the sharpness
parameter $k$ is only weakly constrained, while the activation epoch $a_t$ is
restricted to late times by the requirement of early universe safety.

\subsubsection{Fits}

For the parameter values favored by Fig.~\ref{fig:triangle1}, the best--fit HEED
and $\Lambda$CDM curves in Figs.~\ref{fig:SN}, \ref{fig:CC}, \ref{fig:BAOr}, \ref{fig:BAOt}, \ref{fig:RSD}  are nearly
indistinguishable within the current observational uncertainties. This behavior
is expected.
First, the low--redshift data blocks employed here--SN~Ia distance moduli,
compressed BAO distances, CC $H(z)$ points, and RSD
$f\sigma_8(z)$ measurements---primarily constrain integrated distance and
expansion combinations, and therefore admit substantial degeneracies among
late time expansion histories. In this regime, distinct parameterizations can
map onto very similar $H(z)$ and distance predictions once their parameters are
adjusted to satisfy the same geometric constraints. Second, our SN likelihood
profiles over the nuisance offset $\Delta M$, so SN data constrain relative
distances and provide only weak leverage on the absolute expansion scale.
Finally, in anchored analyses the external Gaussian prior is applied to the
effective late time Hubble scale
\(H_0^{\rm eff}\) rather than directly to $H_0$.
Consequently, the parameters $H_0$ and $c_{e0}^2$ can trade off along the
degeneracy direction implied by the additive HEED embedding while keeping
$H_0^{\rm eff}$ (and hence much of the low--$z$ expansion history) close to the
anchored value. The close overlap of the corresponding best fit curves thus
indicates that HEED can accommodate a nonzero late time entanglement amplitude
without spoiling established low--$z$ distance and growth constraints, while
also highlighting that present data have limited power to distinguish HEED from
$\Lambda$CDM at the level of background and linear growth observables.

\subsubsection{Implications} Taken together, the posterior distributions demonstrate that HEED operates in the regime it was designed for: it activates only at late times, modifies the expansion history at the few percent level, preserves early universe physics, and remains consistent with structure-growth constraints. While the current data do not yet yield a statistically decisive preference over \(\Lambda\)CDM, the results show that HEED can significantly alleviate the Hubble tension without introducing pathologies or fine-tuning, and they motivate further tests with upcoming LSS  and CMB--LSS cross-correlation data. These results should therefore be interpreted as an \emph{anchored consistency test} (high \(H_0^{\rm eff}\) imposed), rather than as a free low-$z$ determination of the Hubble scale.

\section{Conclusions}
\label{sec:conclusions}

We have developed and confronted a simple realization of HEED as a late time, IR
contribution to the cosmic energy budget.
In HEED, the apparent (Hubble) horizon acts as an entanglement screen whose quantum correlations do
not fully saturate the Bekenstein--Hawking entropy at late times. The resulting
deficit is mapped, via horizon thermodynamics and equipartition, into a smooth
bulk component whose characteristic scaling is
\(\rho_{\rm HEED}\propto H^{2}/G\). This scaling is of precisely the magnitude
required to affect late time expansion while remaining negligible at
recombination, and it is qualitatively distinct from free--field vacuum
effects that generate parametrically smaller \(\delta\rho\sim H^{4}\)
corrections.

At the level of internal consistency, HEED is compatible with standard
requirements of cosmological model building. In the additive embedding used
throughout this work, the entanglement sector modifies the background through
an algebraic rescaling of the Friedmann equation while preserving the Bianchi
identity and covariant conservation of matter,
\(\dot\rho_m+3H\rho_m=0\). By construction the model remains physical provided
\(0\le c_e^2(a)<1\), which ensures a positive effective Planck mass
\(M_*^2\propto 1-c_e^2\) and avoids the pathologies associated with vanishing
gravitational coupling. Interpreted as an effective, nonpropagating IR sector,
HEED can transiently realize an effective equation of state below $-1$ during
its switch on without implying ghost instabilities, since no additional
propagating degree of freedom is introduced.

Phenomenologically, HEED leads to several testable late-time signatures. Its
characteristic activation produces a few-percent upward shift of the effective
Hubble scale, modifies distances at \(z\lesssim\mathcal{O}(1)\), and yields a
mild suppression of linear growth, together with a modest enhancement of the
late integrated Sachs--Wolfe (ISW) response. At the same time, the best-fit HEED
and \(\Lambda\)CDM curves for the low-redshift observables considered here
(SN~Ia, BAO, CC, and RSD) are nearly indistinguishable within current
uncertainties. This shows that the HEED sector can be activated at late times
without visibly spoiling established distance and growth constraints, but it
also implies that present low-redshift data do not by themselves provide a
sharp discrimination between the two scenarios. Robust separation will require
higher-precision low-\(z\) measurements, additional growth probes such as weak
lensing and CMB lensing
\cite{Asgari2021KiDS1000,Heymans2021KiDS1000,ACT:2023kun,Carron:2022eyg,HSC:2018mrq},
and possibly CMB--LSS cross-correlations that directly probe the late-time ISW
response through temperature--tracer correlations and related estimators
\cite{Crittenden:1995ak,Afshordi:2003xu,Giannantonio:2008zi,Krolewski:2021znk,Planck:2015fcm,Das:2013sca}.
Although ISW measurements are not included in the present analysis, they provide
an additional in-principle lever arm for distinguishing HEED from purely
geometric late-time modifications.

The present analysis should therefore be interpreted as an anchored
low-redshift consistency test, not as a definitive demonstration that the
Hubble tension is resolved. The likelihood includes an external Gaussian anchor
on the effective late-time Hubble scale \(H_0^{\rm eff}\), so the posterior does
not represent a free inference of a high \(H_0\) from SN+BAO+CC+RSD data alone.
Rather, given a high local expansion scale, we test whether the HEED activation
sector can accommodate it without significantly degrading the fit to current
distance and growth observables. In this sense, the terms ``alleviate'' and
``accommodate'' are used conditionally throughout the paper, and should not be
interpreted as a claim of a statistically established resolution of the Hubble
tension.

Finally, we stress the conceptual advantage of anchoring the IR contribution to
a quasi--local horizon. Teleological issues refer to pathologies that arise when
present time quantities depend on the \emph{entire future history} of the
universe. The future event horizon is a global object: its existence and radius
at time $t$ depend on whether signals emitted after $t$ will ever reach the
observer, which requires knowledge of the spacetime to arbitrarily late times.
Consequently, models that set an IR cutoff by the future event horizon make the
present energy density a functional of the future expansion history, leading to
acausality and nonlocal in--time evolution. In contrast, the apparent (Hubble)
horizon in FLRW is determined quasi--locally on each time slice and underlies
local horizon thermodynamics and derivations of the Friedmann equations via
first--law equipartition arguments. HEED ties its IR contribution to the
apparent (Hubble) horizon and therefore remains local in time, avoiding the
teleology inherent to future event horizon models
\cite{Gibbons1977,Padmanabhan2010RPP,Padmanabhan2012,Wall2018}.

Several extensions are immediate. A next step is a full model comparison with
$\Lambda$CDM using consistent covariances for SN and BAO products and including
additional growth information (cosmic shear, CMB lensing, cluster abundance).
On the theory side, the HEED parameterization can be sharpened by constructing
microphysical models for the deficit function $\delta(a)$ (or $c_e^2(a)$),
e.g.\ in terms of horizon mutual information, entanglement--wedge capacity, or
Gauss law constraints on gravitational dressing. These developments would move
HEED from a phenomenological IR mechanism toward a predictive, first principles
account of how horizon entanglement can influence late time cosmology.

A central limitation of the present analysis is that HEED is formulated here as
an effective FLRW-level description and not yet derived from a fully covariant
microscopic bulk theory. Establishing such an underpinning, or deriving the
activation function from first principles, remains an important problem for
future work. Accordingly, the significance of the present work is not that it supplies a
complete fundamental theory, but that it identifies a concrete IR
mechanism, motivated by horizon thermodynamics, whose background and growth
signatures can be tested directly against cosmological observations.

\medskip

\section*{Acknowledgements}

The work of R.K. and J.S. was partially supported by the Kakos Endowed Chair in Science Fellowship.

\appendix

\section{Entanglement First Law, Modular Energy, and the $H^4$ Scaling}
\label{sec:H4-scaling}

Consider a quantum field theory (beginning with a free/conformal scalar) in $3{+}1$ dimensions on a spatial slice of a FLRW spacetime that is close to de Sitter with Hubble parameter $H$.
Let $B$ be a ball of radius $R$ (possibly of order the Hubble radius) on a constant–time slice.
For a conformal field theory (CFT) in flat space, the vacuum modular Hamiltonian
${\cal H}_B$ for a ball is local and can be written as
\begin{equation}
{\cal H}_B = 2\pi \!\int_B d^3x\, \frac{R^2 - r^2}{2R}\,T_{00}(x) + \text{const.},
\label{eq:modHam}
\end{equation}
as obtained from the conformal map between the ball and a hyperbolic/Rindler
wedge~\cite{Casini2011,BKS2016}.

The first law of entanglement states
that for small state/geometry variations
\begin{equation}
\delta S_B = \delta\langle {\cal H}_B\rangle
= 2\pi \!\int_B d^3x\,\frac{R^2 - r^2}{2R}\,\delta\!\langle T_{00}(x)\rangle,
\label{eq:first-law}
\end{equation}
which is the linearized relation between entanglement entropy and modular energy (relative entropy at first order)~\cite{Bhattacharya2013,Blanco2013}.

In a smooth curved background, the renormalized vacuum stress tensor of a four-dimensional CFT is controlled by the conformal (trace) anomaly, therefore in de Sitter (Weyl-flat with $C_{\mu\nu\rho\sigma}=0$) one has
\begin{equation}
\langle T_{\mu\nu}\rangle_{\rm vac} = c_{\!A}\,H^4\,g_{\mu\nu},
\label{eq:anomaly}
\end{equation}
with $c_{\!A}$ fixed by the type-A anomaly and field content~\cite{Herzog2013}.
Dimensional analysis in $4$D implies $\langle T_{\mu\nu}\rangle \propto H^4$ in the IR.
For a small deformation from Minkowski/adiabatic vacuum to (quasi) de Sitter,
\begin{equation}
\delta\!\langle T_{00}\rangle \sim \sigma\,H^4,
\end{equation}
where $\sigma$ is a dimensionless coefficient set by field content.

Using the spherical weight and
\[
\int_0^R\!4\pi r^2(R^2-r^2)\,dr = \frac{8\pi}{15}R^5,
\]
Eq.~\eqref{eq:first-law} gives
\begin{align}
\delta S_B
&= 2\pi\,\frac{1}{2R}\,\delta\!\langle T_{00}\rangle
\int_B d^3x\,(R^2-r^2) \nonumber\\
&= \frac{8\pi^2}{15}\,R^4\,\delta\!\langle T_{00}\rangle
\sim \frac{8\pi^2}{15}\,R^4\,\sigma\,H^4.
\end{align}

For a ball in a CFT vacuum, the modular (entanglement) temperature is $T_{\rm mod} = 1/(2\pi R)$.
A Clausius-like relation then yields
\begin{align}
\delta E_B &= T_{\rm mod}\,\delta S_B
= \frac{1}{2\pi R} \cdot \frac{8\pi^2}{15}R^4\,\delta\!\langle T_{00}\rangle
\nonumber\\
&= \frac{4\pi}{15}\,R^3\,\delta\!\langle T_{00}\rangle,
\end{align}
and dividing by the ball volume $V_B = (4\pi/3)R^3$ gives the renormalized energy-density shift
\begin{equation}
\delta\rho = \frac{\delta E_B}{V_B} = \frac{1}{5}\,\delta\!\langle T_{00}\rangle
\sim \mathcal{O}(H^4).
\end{equation}

Thus, free/CFT vacuum contributions to the late-time energy density scale as $H^4$ and are negligible compared to the critical density $\rho_{\rm crit} = 3H^2/(8\pi G)$:
\begin{equation}
\frac{\delta\rho}{\rho_{\rm crit}} \sim \left( \frac{H}{M_{\rm Pl}} \right)^2 \ll 1.
\end{equation}

A careful modular Hamiltonian first law analysis shows that standard free-field entanglement yields only
$\mathcal{O}(H^4)$ energy density corrections at late times.
Hence, any phenomenologically relevant $\sim H^2/G$ component must arise from IR/holographic or
nonlocal/emergent physics tied to the horizon scale, not from the UV area law of free fields.

The $H^{4}$ scaling derived above concerns the finite, renormalized, state--dependent
IR response of free/CFT fields to a smooth (quasi--)de Sitter background,
for which the only available macroscopic scale is $H$ and thus $\delta\langle T_{\mu\nu}\rangle\propto H^{4}g_{\mu\nu}$.
By contrast, the UV part of vacuum entanglement across a smooth surface
produces a nonuniversal area--law divergence,
$S_{\rm ent}\sim \eta\,A/\epsilon^{2}$, where $\epsilon$ is a short--distance regulator (UV cutoff)
and $\eta$ depends on field content and the scheme. In semiclassical gravity, the same UV modes
renormalize the Einstein--Hilbert coupling (and higher--curvature terms), so that the geometric
entropy is expressed in terms of the renormalized Newton constant $G_{\rm ren}$ and the UV area term
is not an additional dynamical component~\cite{Jacobson2013,Solodukhin2011,Cooperman2014}.
Counting the area law again as an energy density would therefore double count degrees of freedom
already absorbed into $G_{\rm ren}$. Consequently, free--field entanglement does not generate a
late--time $\mathcal{O}(H^{2}/G)$ contribution: light fields continue to give $\mathcal{O}(H^{4})$ IR effects,
while heavy fields mainly renormalize local couplings.

\section{Data}
\label{data}

\begin{table*}[t]
  \centering
  \caption{CC $H(z)$ measurements (classic 31-point, typically treated as non-correlated). Uncertainties are 1$\sigma$.}
  \label{tab:CC_uncorr_31}
  \begin{ruledtabular}
  \begin{tabular}{ccccc}
    $z$ & $H(z)$ [km s$^{-1}$ Mpc$^{-1}$] & $\sigma$ [km s$^{-1}$ Mpc$^{-1}$] & Refs & Year \\ \hline
    0.07   &  69.0 & 19.6 & \cite{zhang2014four} & 2014 \\
    0.09   &  69.0 & 12.0 & \cite{Simon:2005} & 2005 \\
    0.12   &  68.6 & 26.2 & \cite{zhang2014four} & 2014 \\
    0.17   &  83.0 &  8.0 & \cite{Simon:2005} & 2005 \\
    0.179  &  75.0 &  4.0 & \cite{M_Moresco_2012} & 2012 \\
    0.199  &  75.0 &  5.0 & \cite{M_Moresco_2012} & 2012 \\
    0.200  &  72.9 & 29.6 & \cite{zhang2014four} & 2014 \\
    0.270  &  77.0 & 14.0 & \cite{Simon:2005} & 2005 \\
    0.280  &  88.8 & 36.6 & \cite{zhang2014four} & 2014 \\
    0.352  &  83.0 & 14.0 & \cite{M_Moresco_2012} & 2012 \\
    0.3802 &  83.0 & 13.5 & \cite{Moresco_2016} & 2016 \\
    0.400  &  95.0 & 17.0 & \cite{Simon:2005} & 2005 \\
    0.4004 &  77.0 & 10.2 & \cite{Moresco_2016} & 2016 \\
    0.4247 &  87.1 & 11.2 & \cite{Moresco_2016} & 2016 \\
    0.4497 &  92.8 & 12.9 & \cite{Moresco_2016} & 2016 \\
    0.470  &  89.0 & 34.0 & \cite{Ratsimbazafy_2017} & 2017 \\
    0.4783 &  80.9 &  9.0 & \cite{Moresco_2016} & 2016 \\
    0.480  &  97.0 & 62.0 & \cite{Daniel_Stern_2010} & 2010 \\
    0.593  & 104.0 & 13.0 & \cite{M_Moresco_2012} & 2012 \\
    0.680  &  92.0 &  8.0 & \cite{M_Moresco_2012} & 2012 \\
    0.781  & 105.0 & 12.0 & \cite{M_Moresco_2012} & 2012 \\
    0.875  & 125.0 & 17.0 & \cite{M_Moresco_2012} & 2012 \\
    0.880  &  90.0 & 40.0 & \cite{Daniel_Stern_2010} & 2010 \\
    0.900  & 117.0 & 23.0 & \cite{Simon:2005} & 2005 \\
    1.037  & 154.0 & 20.0 & \cite{M_Moresco_2012} & 2012 \\
    1.300  & 168.0 & 17.0 & \cite{Simon:2005} & 2005 \\
    1.363  & 160.0 & 33.6 & \cite{MorescoMNRAS2015} & 2015 \\
    1.430  & 177.0 & 18.0 & \cite{Simon:2005} & 2005 \\
    1.530  & 140.0 & 14.0 & \cite{Simon:2005} & 2005 \\
    1.750  & 202.0 & 40.0 & \cite{Simon:2005} & 2005 \\
    1.965  & 186.5 & 50.4 & \cite{MorescoMNRAS2015} & 2015 \\
  \end{tabular}
  \end{ruledtabular}
\end{table*}

\begin{table}[t]
\caption{\label{tab:BAO_combined}
BAO transverse comoving and radial (Hubble) distance measurements, shown jointly as $D_M(z)/r_d$ and $D_H(z)/r_d$ with $1\sigma$ uncertainties. Numerical values are taken from Table~4 of Ref.~\cite{1607.03155}, Table~III of Ref.~\cite{Alam2021eBOSS} and Table~IV of Ref.~\cite{DESI2025DR2_BAO}. BOSS/eBOSS BAO data are also summarized in Ref.~\cite{SDSSIV_Final_BAO_RSD}.}
\begin{ruledtabular}
\begin{tabular}{ccccccc}
$z$ & $D_M/r_d$ & $\sigma_{D_M/r_d}$ & $D_H/r_d$ & $\sigma_{D_H/r_d}$ & Refs & Year \\ \hline
0.38  & 10.23   & 0.17  & 25.00   & 0.76  & \cite{Alam2021eBOSS} & 2020 \\
0.38  & 10.16   & 0.18  & 24.59   & 0.72  & \cite{1607.03155}          & 2016 \\
0.51  & 13.588  & 0.167 & 21.863  & 0.425 & \cite{DESI2025DR2_BAO}          & 2025 \\
0.51  & 13.36   & 0.21  & 22.33   & 0.58  & \cite{Alam2021eBOSS} & 2020 \\
0.51  & 13.6    & 0.2   & 22.49   & 0.62  & \cite{1607.03155}          & 2016 \\
0.70  & 17.86   & 0.33  & 19.33   & 0.53  & \cite{Alam2021eBOSS} & 2020 \\
0.706 & 17.351  & 0.177 & 19.455  & 0.33  & \cite{DESI2025DR2_BAO}          & 2025 \\
1.321 & 27.601  & 0.318 & 14.176  & 0.221 & \cite{DESI2025DR2_BAO}          & 2025 \\
1.48  & 30.69   & 0.80  & 13.26   & 0.55  & \cite{Alam2021eBOSS} & 2020 \\
1.484 & 30.512  & 0.76  & 12.817  & 0.516 & \cite{DESI2025DR2_BAO}          & 2025 \\
2.33  & 38.988  & 0.531 & 8.632   & 0.101 & \cite{DESI2025DR2_BAO}          & 2025 \\
2.33  & 37.6    & 1.9   & 8.93    & 0.28  & \cite{Alam2021eBOSS} & 2020 \\
\end{tabular}
\end{ruledtabular}
\end{table}

\begin{table}[t]
\caption{\label{tab:fs8_tableI}
RSD growth rate measurements $f\sigma_8(z)$ and 1$\sigma$ uncertainties.
Data acquired from Table II of Ref.~\cite{RSDcompile63}.
}
\begin{ruledtabular}
\begin{tabular}{cccccc}
$\text{Index}$ & $\text{Dataset}$ & $z$ & $f\sigma_8(z)$ & $\text{Refs}$ & $\text{Year}$ \\ \hline
1  & SDSS-LRG        & 0.35  & $0.440 \pm 0.050$   & \cite{JCAP.0910.004} & 2006 \\
2  & VVDS            & 0.77  & $0.490 \pm 0.180$   & \cite{JCAP.0910.004} & 2009 \\
3  & 2dFGRS          & 0.17  & $0.510 \pm 0.060$   & \cite{JCAP.0910.004} & 2009 \\
4  & 2MRS            & 0.02  & $0.314 \pm 0.048$   & \cite{1011.3114}, \cite{1203.4814} & 2010 \\
5  & SnIa+IRAS       & 0.02  & $0.398 \pm 0.065$   & \cite{1203.4814}, \cite{1111.0631} & 2011 \\
6  & SDSS-LRG-200    & 0.25  & $0.3512 \pm 0.0583$ & \cite{1102.1014} & 2011 \\
7  & SDSS-LRG-200    & 0.37  & $0.4602 \pm 0.0378$ & \cite{1102.1014} & 2011 \\
8  & SDSS-LRG-60     & 0.25  & $0.3665 \pm 0.0601$ & \cite{1102.1014} & 2011 \\
9  & SDSS-LRG-60     & 0.37  & $0.4031 \pm 0.0586$ & \cite{1102.1014} & 2011 \\
10 & WiggleZ         & 0.44  & $0.413 \pm 0.080$   & \cite{1204.3674} & 2012 \\
11 & WiggleZ         & 0.60  & $0.390 \pm 0.063$   & \cite{1204.3674} & 2012 \\
12 & WiggleZ         & 0.73  & $0.437 \pm 0.072$   & \cite{1204.3674} & 2012 \\
13 & 6dFGS           & 0.067 & $0.423 \pm 0.055$   & \cite{1204.4725} & 2012 \\
14 & SDSS-BOSS       & 0.30  & $0.407 \pm 0.055$   & \cite{1203.6565} & 2012 \\
15 & SDSS-BOSS       & 0.40  & $0.419 \pm 0.041$   & \cite{1203.6565} & 2012 \\
16 & SDSS-BOSS       & 0.50  & $0.427 \pm 0.043$   & \cite{1203.6565} & 2012 \\
17 & SDSS-BOSS       & 0.60  & $0.433 \pm 0.067$   & \cite{1203.6565} & 2012 \\
18 & Vipers          & 0.80  & $0.470 \pm 0.080$   & \cite{1303.2622} & 2013 \\
19 & SDSS-DR7-LRG    & 0.35  & $0.429 \pm 0.089$   & \cite{1209.0210} & 2013 \\
20 & GAMA            & 0.18  & $0.360 \pm 0.090$   & \cite{1309.5556} & 2013 \\
21 & GAMA            & 0.38  & $0.440 \pm 0.060$   & \cite{1309.5556} & 2013 \\
22 & BOSS-LOWZ       & 0.32  & $0.384 \pm 0.095$   & \cite{1312.4854} & 2013 \\
23 & SDSS DR10 and DR11  & 0.32  & $0.480 \pm 0.100$ & \cite{1312.4854} & 2013 \\
24 & SDSS DR10 and DR11  & 0.57  & $0.417 \pm 0.045$ & \cite{1312.4854} & 2013 \\
25 & SDSS-MGS        & 0.15  & $0.490 \pm 0.145$   & \cite{1409.3238} & 2015 \\
26 & SDSS-veloc      & 0.10  & $0.370 \pm 0.130$   & \cite{1503.05945} & 2015 \\
27 & FastSound       & 1.40  & $0.482 \pm 0.116$   & \cite{1511.08083} & 2015 \\
28 & SDSS-CMASS      & 0.59  & $0.488 \pm 0.060$   & \cite{1312.4889} & 2016 \\
29 & BOSS DR12       & 0.38  & $0.497 \pm 0.045$   & \cite{1607.03155} & 2016 \\
30 & BOSS DR12       & 0.51  & $0.458 \pm 0.038$   & \cite{1607.03155} & 2016 \\
31 & BOSS DR12       & 0.61  & $0.436 \pm 0.034$   & \cite{1607.03155} & 2016 \\
32 & BOSS DR12       & 0.38  & $0.477 \pm 0.051$   & \cite{1607.03150} & 2016 \\
33 & BOSS DR12       & 0.51  & $0.453 \pm 0.050$   & \cite{1607.03150} & 2016 \\
34 & BOSS DR12       & 0.61  & $0.410 \pm 0.044$   & \cite{1607.03150} & 2016 \\
35 & Vipers v7       & 0.76  & $0.440 \pm 0.040$   & \cite{1610.08362} & 2016 \\
36 & Vipers v7       & 1.05  & $0.280 \pm 0.080$   & \cite{1610.08362} & 2016 \\
37 & BOSS LOWZ       & 0.32  & $0.427 \pm 0.056$   & \cite{1606.00439} & 2016 \\
38 & BOSS CMASS      & 0.57  & $0.426 \pm 0.029$   & \cite{1606.00439} & 2016 \\
39 & Vipers          & 0.727 & $0.296 \pm 0.0765$  & \cite{1611.07046} & 2016 \\
40 & 6dFGS+SnIa      & 0.02  & $0.428 \pm 0.0465$  & \cite{1611.09862} & 2016 \\
41 & Vipers          & 0.60  & $0.480 \pm 0.120$   & \cite{1612.05647} & 2016 \\
42 & Vipers          & 0.86  & $0.480 \pm 0.100$   & \cite{1612.05647} & 2016 \\
43 & Vipers PDR-2    & 0.60  & $0.550 \pm 0.120$   & \cite{1612.05645} & 2016 \\
44 & Vipers PDR-2    & 0.86  & $0.400 \pm 0.110$   & \cite{1612.05645} & 2016 \\
45 & SDSS DR13       & 0.10  & $0.480 \pm 0.160$   & \cite{1612.07809} & 2016 \\
46 & 2MTF            & 0.001 & $0.505 \pm 0.085$   & \cite{1706.05130} & 2017 \\
47 & Vipers PDR-2    & 0.85  & $0.450 \pm 0.110$   & \cite{1708.00026} & 2017 \\
48 & BOSS DR12       & 0.31  & $0.469 \pm 0.098$   & \cite{1709.05173} & 2017 \\
49 & BOSS DR12       & 0.36  & $0.474 \pm 0.097$   & \cite{1709.05173} & 2017 \\
50 & BOSS DR12       & 0.40  & $0.473 \pm 0.086$   & \cite{1709.05173} & 2017 \\
51 & BOSS DR12       & 0.44  & $0.481 \pm 0.076$   & \cite{1709.05173} & 2017 \\
52 & BOSS DR12       & 0.48  & $0.482 \pm 0.067$   & \cite{1709.05173} & 2017 \\
53 & BOSS DR12       & 0.52  & $0.488 \pm 0.065$   & \cite{1709.05173} & 2017 \\
54 & BOSS DR12       & 0.56  & $0.482 \pm 0.067$   & \cite{1709.05173} & 2017 \\
55 & BOSS DR12       & 0.59  & $0.481 \pm 0.066$   & \cite{1709.05173} & 2017 \\
56 & BOSS DR12       & 0.64  & $0.486 \pm 0.070$   & \cite{1709.05173} & 2017 \\
57 & SDSS DR7        & 0.10  & $0.376 \pm 0.038$   & \cite{1712.04163} & 2017 \\
58 & SDSS-IV         & 1.52  & $0.420 \pm 0.076$   & \cite{1801.02689} & 2018 \\
59 & SDSS-IV         & 1.52  & $0.396 \pm 0.079$   & \cite{1801.02656} & 2018 \\
60 & SDSS-IV         & 0.978 & $0.379 \pm 0.176$   & \cite{1801.03043} & 2018 \\
61 & SDSS-IV         & 1.23  & $0.385 \pm 0.099$   & \cite{1801.03043} & 2018 \\
62 & SDSS-IV         & 1.526 & $0.342 \pm 0.070$   & \cite{1801.03043} & 2018 \\
63 & SDSS-IV         & 1.944 & $0.364 \pm 0.106$   & \cite{1801.03043} & 2018 \\
\end{tabular}
\end{ruledtabular}
\end{table}

\newpage


\bibliography{main.bbl}

\end{document}